\documentclass[12pt,preprint]{aastex}

\newcommand{\etal}{ et al. }
\newcommand{\simgt}{\ga}
\newcommand{\simlt}{\lower.5ex\hbox{$\; \buildrel < \over \sim \;$}}

\slugcomment{draft date: \today}

\begin{document}

\title{Shocked Molecular Gas in the Supernova Remnants W~28 and W~44:
Near-infrared and millimeter-wave observations}

\author{William T. Reach, Jeonghee Rho, and T. H. Jarrett}

\affil{Infrared Processing and Analysis Center, 
California Institute of Technology,
Pasadena, CA 91125}

\email{reach@ipac.caltech.edu}

\def\coone{ CO($1\rightarrow 0$)}
\def\cotwo{ CO($2\rightarrow 1$)}
\def\cothree{ CO($3\rightarrow 2$)}
\def\cstwo{ CS($2\rightarrow 1$)}
\def\csthree{ CS($3\rightarrow 2$)}
\def\csfive{ CS($5\rightarrow 4$)}
\def\hcop{ HCO$^+$($1\rightarrow 0$)}
\def\cothirteen{ $^{13}$CO($1\rightarrow 0$)}

\begin{abstract}
High resolution millimeter-wave (CO, CS, and HCO$^+$ rotational lines) and
near-infrared (H$_2$ 2.12 $\mu$m rovibrational line, \ion{Fe}{2} 
fine-structure line) observations of
the supernova remnants W~28 and W~44 reveal extensive shocked molecular
gas where supernova blast waves are propagating  into giant
molecular clouds. New CO observations were carried out with the IRAM 30-m and
ARO 12-m telescopes, and the near-infrared observations were with Prime
Focus Infrared Camera and Wide-Field Infrared Camera 
on the Palomar Hale 200-inch telescope.  
The near-infrared observations reveal shocked H$_2$
emission from both supernova remnants, 
showing intricate networks of filaments
on arcsec scales, following the bright ridges of  the  radio shells.
The emission is particularly bright in the northeastern, southern
and western parts of W44 and the eastern bar in W28.  The  H$_2$ emission
reveals some bright, clumpy structures as well as very thin
filamentary structures likely to be individual shock fronts seen edge-on.   
The high-resolution IRAM CO(2-1) and
CS(2-1) spectra clearly distinguish between the shocked and
pre-shock  gas for most of SNRs. Some of the CO spectra appear to
have multiple components,
but the less-optically-thick $^{13}$CO lines clearly demonstrate
that the CO(2-1) lines are broad, with deep absorption dips caused by 
cold, dense gas in the light of sight. 
The CO and CS linewidths, indicative of the shock
speed, are $20$--$30$ km~s$^{-1}$.

Both the near-infrared and millimeter-wave emission are attributed to
shocks into gas with density $>10^3$ cm$^{-3}$. Individual
shock structures are resolved in the H$_2$ emission, with inferred
edge-on shock thickness $\sim 10^{17}$ cm, consistent with
non-dissociative shocks into gas densities of $10^{3}$ - $10^{4}$
cm$^{-3}$. Bright 1720 MHz OH masers are located within the shocked
H$_2$ gas  complexes and highlight only localized areas where the
conditions for masing are optimal. The H$\alpha$ and X-ray emission,
which trace hotter shocked gas, have morphologies very different
from the radio. We find a detailed correlation of the radio and H$_2$
emission for some long filaments, indicating cosmic ray acceleration or
re-acceleration due to the shocks into moderately dense gas. Compared
to the interclump gas and the very dense cores, the synchrotron
emissivity of the moderate-density (CO-emitting) medium
is highest, which explains the
radio-H$_2$ correlation and  the  very bright radio emission 
of these two SNRs despite their relatively advanced age. 
The different morphologies of these 
two remnants at different wavelengths is explained by a highly
nonuniform structure for giant molecular clouds, with low-density
($\sim 5$ cm$^{-3}$) gas occupying most ($\sim 90$\%) of the volume,
moderate density gas ($\sim 10^3$ cm$^{-3}$) gas occupying most of the
rest of the volume, and dense gas in cores.

{\it This preprint is made available through astro-ph and has had its
figures compressed significantly from the originals. Please see the
published article (expected in Jan 2005 in the Astrophysical Journal)
for higher-quality figures.}
\end{abstract}
                            
\section{Introduction}

W~28 and W~44 are two supernova remnants in molecular clouds. 
They are very bright radio sources---the 6$^{th}$ and
7$^{th}$ brightest remnants in the \citet{green} catalog---with 
shell-like morphologies \citep{W28radiodubner,W44radio}, located adjacent to
giant molecular clouds with which they have been suspected to
be interacting \citep{wootten,denoyer}. 
W~28 and W~44 are also paradigms of the new class of `mixed-morphology'
supernova remnants, whose shell-like, non-thermal radio emission
contrasts sharply with centrally-filled, thermal X-ray emission
\citep{long91,RP98}. Both W~28 and W~44 also contain OH 1720 MHz masers
attributed to dense, shocked gas \citep{lockett}.
Based on absorption of the radio continuum by foreground gas,
the distances of W~28 and W~44 are estimated to be
1.9 kpc \citep{velazquez} and 2.5 kpc \citep{cox99}, respectively.
Both remnants have mean radii of 11 pc.
The radio shell of W~44 is non-circular, being elongated north-south and
brighter on the eastern side than elsewhere. 
The radio shell of W~28 is much brighter and better defined on its northern
side than elsewhere.

W~44 is host to the pulsar B1853+01, which is almost certainly the
remnant of the progenitor star \citep{wolzscan}. 
The presence of the pulsar yields
two key clues to the nature of W~44. First, it must be the
result of a core-collapse supernova from a progenitor with
mass between 8 and 20 $M_\odot$:
a smaller progenitor would not form a neutron star 
\citep{woosley,wheeler81}, 
while a larger one would form a black hole instead \citep{fryer}.
The spectral type of the progenitor, when on the main sequence, 
would have been between B4 and O8.
If such a star were surrounded by
material with a density 1 or $10^3$ cm$^{-3}$, the \ion{H}{2} region 
would have been smaller than 2 or 0.02 pc, respectively.
Thus  the progenitor would have
had little influence on its parent molecular cloud on the size
scales of the present supernova remnant
\citep{spitzer}.
Second, the age of the remnant is probably comparable 
to the spin-down age of
the pulsar, $2\times 10^4$ yr \citep{wolzscan}.
PSR B1853+01 also produces relativistic particles and excites
a wind nebula visible in the radio \citep{frail44} and
X-ray \citep{petre44}.
A relatively young pulsar, PSR B1758-23, is near W~28 but outside
the remnant \citep{kaspi28}; the weight of current evidence 
suggests that this pulsar is not related to W~28 \citep{claussen28}.
The association of W~28 with molecular clouds and its radio/X-ray
morphology suggest the progenitor may also have been a core-collapse,
in which case the stellar remnant is now a radio-quiet neutron star
or a black hole; however, the supernova could also have been a Type I,
coincidentally close to a molecular cloud.

$\gamma$-ray sources, detected with the {\it Compton}
Gamma Ray Observatory using EGRET, have been associated with W~28
and W~44 \citep{esposito}; the updated associations using 
the Third EGRET source catalog \citep{hartman} 
are 3EG J1800-2338 for W~28 and 3EG J1856+0114 for W~44.
The angular resolution in $\gamma$-rays
is insufficient ($0.3^\circ$) to pinpoint their origin. 
The W~28 source is south of remnant center in a relatively radio-faint 
part of the remnant. 
The W~44 source is located within the remnant, with the pulsar
included in its error circle.
No TeV photons were detected from W~28 using
CANGAROO, so the source spectrum must turn off between GeV and 
TeV energies \citep{rowell}.

Both W~28 and W~44 have extensive evidence for interaction with molecular
clouds. Early observations \citep{wootten,denoyer} showed molecular gas
near or possibly in the remnant, but the sensitivity, resolution and
available millimeter-wave receivers (limited to the 3mm band) were
inadequate to clearly reveal a direct interaction with the molecular
clouds. X-ray observations showed that these two remnants are filled
with a large amount of relatively dense ($n\sim 1$ cm$^{-3}$), hot gas. 
The X-ray morphology contrasts markedly from the
shell-like radio morphology, making W~28 and W~44 members of 
the ``mixed-morphology'' class; the existence of a significant
amount of material inside the remnant was attributed to interaction
of the remnant with relatively dense gas \citep{rho44,RP98}. 

The most clear-cut evidence for interaction between the supernova
remnants and molecular clouds is the detection of emission from the
shocked molecules themselves. 
The molecules can be directly detected in four ways: millimeter-wave
emission lines with linewidths much greater than those of cold gas,
OH 1720 MHz maser emission, far-infrared line emission, and 
near-infrared H$_2$ emission.
(1) Broad molecular line emission has clearly been detected from 
W~44 using \coone\ and \cotwo\ observations \citep{setaw44}, and from W~28
using a small telescope to map the remnant in the \cothree\ line
\citep{arikawa}. The linewidths $\sim 20-30$ km~s$^{-1}$ FWHM,
with maximum extents up to 70 km~s$^{-1}$ in some locations, 
clearly distinguish the shocked gas from the cold, ambient gas.
(2) OH 1720 MHz maser emission has been detected from
W~28 \citep{frail94} and observed in more detail for both W~28 and W~44
\citep{claussen} using the NRAO Very Large Array.
These 1720 MHz OH masers have been interpreted as spots of
amplified radio emission through clumps of OH gas with densities
and temperatures characteristic of modest-density ($n\sim 10^3$ cm$^{-3}$)
gas that has been shocked by modest-velocity ($v\sim 20$ km~s$^{-1}$)
non-dissociative shocks \citep{lockett,wardle}. 
The brightest maser emission arises from small spots with high
amplification. VLBI observations yield upper limits to the
angular sizes from 0.05--0.18$''$, with very strong measured magnetic
field strengths of 2 mG \citep{claussenmerlin}.
Recent radio spectral observations of W~28
showed that the normal, thermal absorption in the main OH lines 
can trace the shocked gas, because
the absorption lines have broad line widths; faint satellite
line emission (probably weak masers), has a narrow line width and covers much
more area than just the maser peaks \citep{yusefzadeh03a}.
There now appears to be a strong association between `mixed morphology'
supernova remnants and maser-emitting supernova remnants
\citep{yusefzadeh03}, suggesting that while W~28, W~44, and IC~443 may
be prototypes of molecular-cloud-interacting supernova remnants,
the total number could be much higher.
(3) Highly-excited far-infrared CO($16\rightarrow 15$) 
and mid-infrared H$_2$ (S3 and S9) emission,
together with all of the atomic fine structure lines expected from
shocks into moderate-density gas,
were detected from W~28 and W~44
using the {\it Infrared Space Observatory}
\citep{rr98,rr00}. 
(4) Near-infrared H$_2$ emission is
predicted to be one of the main coolants of shocks in
dense gas \citep{DRD}. Bright H$_2$ 2.12 $\mu$m emission has
been detected from IC~443 \citep{ic443h2,rho443}, and 3C~391 \citep{rr02}.
Near-infrared emission from W~28 and W~44
has not yet been reported, and one of the main goals of
this paper is to present our new observations and their
implications. 

Some relevant theoretical models have recently been published.
A model was developed specifically for W~44, attempting to explain
its observed features across an exceptionally wide range of 
observed properties, analytically by \citet{cox99} and 
numerically by \citet{shelton}.
They explain many properties of the remnant
as the result of a shock propagating into an essentially uniform
medium with a density of $n_0=6$ cm$^{-3}$.
The hot interior of the
remnant is relatively dense in their model, $n_{int}=1$ cm$^{-3}$,
as required to explain the centrally-filled X-rays.
This model does not include denser gas.
\citet{chevalier} published a model of a supernova remnant in a
molecular cloud. This model is generally similar to that 
of \citet{cox99}, because most of the observable properties
are due to shocks in the interclump gas with a pre-shock density 
$n_0=4$--5 cm$^{-3}$; however, \citet{chevalier} also 
discussed some of the consequences of denser material.
Based on these models, one would infer that some giant molecular
clouds are predominantly
composed of material of much lower density than is
capable of producing the features that define giant molecular
clouds, such as CO emission, gravitational
binding, and star formation.
Low-density ($n<10$ cm$^{-3}$) material is either diffuse 
(tenuous, transient, and unbound) or is present together with 
other material that 
we know of as the molecular cloud and dominates the mass and
most other observable properties.
How can these two different views of molecular clouds be reconciled?

The observations of W~28 and W~44 provide a special opportunity to
determine the structure of giant molecular clouds by studying
the results of $10^{51}$ erg explosions inside the clouds. 
By observing the results of a supernova explosion 
in the cloud, we can determine whether 
(1) the shock expands unimpeded, as would occur if the
filling factor of dense gas is very small and there is no
interclump gas; or
(2) the shocks propagate primarily in the dense gas, 
as would occur if its filling factor is large; or
(3) the shock is not significantly affected by the dense
gas (having low filling factor) but propagates primarily into
low-density, interclump gas.
A wide range of supernova remnant morphologies is possible,
and at least three pre-shock density regimes have been
inferred from the range of gas coolants in far-infrared spectra 
\citep{rr00}.
By observing the shocked molecular gas directly, and at unprecedented 
sensitivity and resolution, we can better explain the
current remnant morphology and begin to attribute different
aspects of the supernova remnants to specific properties of the 
parent molecular clouds.

After this Introduction, we summarize the new observations that we performed:
wide-area millimeter wave CO observations, high-resolution
millimeter-wave CO and CS observations, and near-infrared H$_2$ observations.
Then we describe the results for each remnant in detail.
With all of the observations and results in hand, we then
compare the shocks as traced by 
near-infrared and millimeter-wave emission and attempt to separate 
pre-shock gas, gas experiencing dissociative shocks, and
gas experiencing non-dissociative shocks. 
We discuss the implications of our observations for the origin of
synchrotron radiation from mature supernova remnants and for
cosmic ray acceleration.
Finally, we explain why the properties of the ambient 
medium that we infer from infrared and millimeter observation
of shocked clumps and filaments is 
so different from the properties inferred from X-ray and radio 
observations of the supernova shell and hot interior.

\section{Observations}

\subsection{CO Survey data}

The parent clouds with which the remnants are interacting, and from which 
the progenitors presumably formed, can be seen in CO surveys of the
galactic plane. 
Previous studies of these remnants have referred to a wide range of
structures in the interstellar medium as being associated with
the remnants, including \ion{H}{1} emission and absorption
\citep{knapp74,denoyer,velazquez} and limited CO mapping
\citep{dickel76,wootten77,wootten,setaw44}
which trace parts of the clouds but don't reveal their full extents.

\begin{figure}[th]
\epsscale{1.5}
\epsscale{1}
\figcaption[f1a.eps]{
{\bf FIGURE NOT INCLUDED IN TEX FILE. SEE FILE 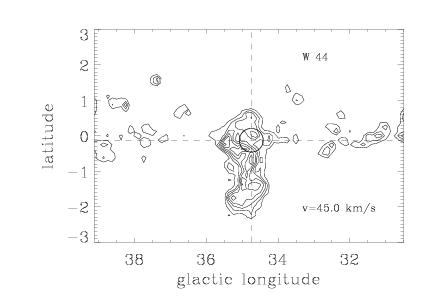 and 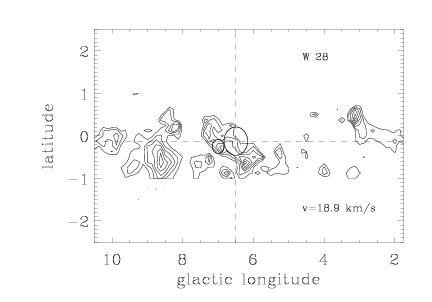.}
CO images of the giant molecular clouds
containing the supernova remnants W~44 and W~28.
Contours are at constant CO(1-0) antenna temperature, ranging from 
1--30 K (0.65 km~s$^{-1}$ channel) and 3--12 K (1.3 km~s$^{-1}$ channel)
for W~44 and W~28, respectively.
The thick circle with dashed cross-hairs through the center indicates
the size and center of the radio emission from each remnant. A smaller
circle, to the left of the W~28, indicates the location of the Trifid 
Nebula.
\label{comini}}
\end{figure}

Figure~\ref{comini} shows the CO images of the parent molecular clouds.
The CO data are from the reprocessed survey \citet{dht}, including
the inner galaxy data originally from \citet{bitran}.
W~44 is located in a well-defined, giant molecular cloud 
that is clearly distinguished from surrounding material; the cloud
is centered at $l=35.0^\circ$, $b=-0.7^\circ$, $v=44$ km~s$^{-1}$, with
a radius of 58 pc and a total mass of $1.8\times 10^6$ $M_\odot$
\citep{dame86}.
The parent cloud for W~28 can be discerned in the
CO data-cube for the inner galaxy \citep{dht}; it is centered at
$l=6.6^\circ$, $b=0.0^\circ$, $v=19$ km~s$^{-1}$, with
a diameter $\sim 25$ pc and a mass of $1.4\times 10^6$ $M_\odot$.
The similar radio and X-ray morphologies, ages, locations near
molecular clouds, and parent molecular cloud properties make
W~44 and W~28 a good pair of remnants to study together.

\subsection{NRAO/Steward Observatory 12-meter Observations}

Observations were performed in March 1997 and
November 2002 with the 12-m millimeter-wave telescope on Kitt Peak.
The observations were made using the On-The-Fly technique, whereby the telescope
was slewed in right ascension at a constant rate of 30 and 10 $''$/s,
and spectra were taken at 0.1 sec intervals, 
which finely samples the telescope beam of 53$''$ and 26$''$
at 115 and 230 GHz, respectively.
In March 1997, the eastern half of W~44 was mapped in the \coone\ line 
(115 GHz), and regions surrounding some of the apparent shock interactions 
in W~44 and W~28 were mapped in the \cotwo\ line (230 GHz). In November 2002,
a complete map of W~44 was made in the \cotwo\ line. 

The \coone\ observations of W~44 reveal extensive emission covering the entire region.
On large scales, the dominant feature is the giant molecular cloud (GMC) at 
a velocity of 43 km/s with lines that are typically 10 km~s$^{-1}$ wide
(with substructure); this is consistent with the overall giant molecular cloud
described above. 
There were no clear 
indications of shock-broadened molecular lines from the \coone\ observations;
most of the broadening is due to the same combination of gravity, turbulence,
and magnetic pressure that generates the linewidths observed from this and
other GMCs. 

The \cotwo\ observations of both W~28 and W~44 also reveal extensive emission, 
with a great deal of structure both spatially and spectrally. 
The separation of
the emission into parent cloud, shocked gas, and foreground absorption requires
understanding the `geography' of each remnant, which we describe in \S~\ref{w44sec}
and~\ref{w28sec} for W~44 and W~28, respectively.

\subsection{IRAM 30-meter Observations}

On September 16-22, 1997, we observed
W~44 and W~28 using the IRAM 30-m telescope on Pico Veleta, Spain.
Some spectra were also obtained from our earlier November 22-29, 1996 
observing run.
These observations cover small portions of the remnants, selected based on
the 12-m maps and the \coone\ maps of \citet{setaw44}.
The observing procedure, calibration, and data reduction for these
observations were the same as described for our previous 
observations \citep{rr99}.
Three receivers were used to simultaneously observe three different spectral 
lines at 1.3, 2, and 3 mm.
The weather during the 1997 observing run included freezing rain, but
had usable periods of haze. The receivers were tuned to 
the \cstwo, \csthree, and \cotwo\ lines, with system temperatures 
typically 330, 520, and 1300 K. Very few lines of sight produced
detectable \cstwo\, so we tuned the 3 mm receiver
to SiO($2\rightarrow 1$, v=0) for the last half of the run.
The weather during the 1996 run was better, which allowed 
\csthree\ and \csfive\ observations.

\subsection{Palomar Observations}

On July 12-13, 2001, and August 16-17, 2002, we observed portions of the W~44 
and W~28 (and other supernova remnants) using the Prime Focus Infrared Camera 
(PFIRCAM) on the Hale 200-inch 
telescope on Mount Palomar. The PFIRCAM has a $256\times 256$ pixel
array, with a pixel scale of 0.494$''$ at the f/3.3 prime focus of 
the 200-inch telescope. In 2001, the weather was hazy, with 1.5$''$ seeing.
In 2002, the weather was very good, with 0.8$''$ seeing 
on both nights. 
Two fields (the western portion of W~44, and the eastern portion 
of W~28) were re-observed using the new 
Wide-Field Infrared Camera (WIRC), a 2048$\times$2048 camera with 
0.25$''$ pixels, on August 9, 2003, also under 
very good conditions (0.9$''$).
Our observing strategy, calibration,
and data reduction were the same as for our previous observations 
\citep{rr02}.
In summary, we made dithered rasters switching between the remnant
and a reference position outside the remnant, combined the reference
observations to make a sky image, and combined sky-subtracted images
on the remnant into mosaics.
The data reduction procedure was improved to take
full advantage of 2MASS \citep{twomassref} astrometry, by locating 
cataloged stars on each image, determining an accurate world coordinate
system for each image, then generating the combined mosaic.
 The calibration procedure was refined to
use improved 2MASS magnitudes from recent re-processing, and we restricted
the range of magnitudes for calibrators to be well within the range of
high-confidence detection 2MASS and linearity in PFIRCAM. 

All regions were
observed through the H$_2$ 2.12 $\mu$m filter, 
which is 1\% wide in PFIRCAM and 1.5\% in WIRC.
Smaller regions were observed in other narrow (1\% wide) filters to measure
the continuum at 2.2 $\mu$m, the [\ion{fe}{2}] 1.64 $\mu$m line
and nearby continuum, and the Paschen $\beta$ 1.28 $\mu$m line.
Extended emission was present in both the H$_2$ and [\ion{fe}{2}] 
filters for W~44 and W~28. No extended emission was detected in the
continuum filters or P$\beta$ ($< 10^{-5}$ erg~s$^{-1}$~cm$^{-2}$~sr$^{-1}$). 
Therefore, continuum subtraction is neither
needed nor performed on the observations presented below. 
Some artifacts remain in the PFIRCAM images, specifically: faint ghost images
located 13$''$ NW (position angle -61$^\circ$)
from very bright sources (nearly impossible to discern in
the published images because of the huge number of real stars that
are brighter than the ghost images); and rays of stray light when
stars fall on the edge of the detector (some were masked by hand and
appear as polygonal cutouts). The observations have a noise level
of 0.5--$1\times 10^{-5}$ erg~s$^{-1}$~cm$^{-2}$~sr$^{-1}$ per pixel.
The absolute calibration is accurate to better than 20\%, and the
narrow-band directly measure the surface brightness of the shocked gas
except in positions containing stars.


\subsection{Other observational data}

To complement the new CO and molecular data, we requested radio and optical
images from other researchers. Supernova remnants are most clearly traced by 
nonthermal radio continuum emission, so the best maps of overall remnant 
geography are low-frequency but high-resolution radio images.
For W~44, we used the VLA image of \citet{W44radio} at a frequency of
1465 MHz; the $15''$ angular resolution and low (1.5 mJy/beam) noise
of this image clearly reveals the 
filamentary internal structure of the remnant. The astrometry was corrected
by comparing to the high-resolution radio image of the W~44 pulsar wind nebula
published by \citet{frail44}, using the pulsar as a guide point.
For W~28, we used the VLA image of \citet{W28radiodubner} at a frequency of 
328 MHz; the $92''\times 52''$ angular resolution is not good but the
remnant is cleanly separated from thermal emission. We compared this image
to the portion of the \citet{W28radiofrail} image that covered W~28
serendipitously and very found good correspondence of the remnant features, 
as expected. 

H$\alpha$ images of W~28 and W~44 were created by combining digitized,
deep (3 hr) plates
taken at the 48-inch UK Schmidt telescope and provided as part of the
SuperCOSMOS H$\alpha$ Survey \citep{parker,malin,macgillivray}. 
The line filter has a 1\% width centered on 6590 \AA.
Images through the narrow continuum filter show no extended emission, so
we did not subtract them from the H$\alpha$ images used here.
To verify the astrometry, we compared the reprojected, combined H$\alpha$ 
mosaic to the Digitized Sky Survey and HD catalog stars;
the DSS and HD positions were within $1''$ while there was an
offset of $9''$ in the H$\alpha$ image, which we corrected.
The H$\alpha$ emission from W~28 has been described before 
by \citet{vandenbergh}, but the new image is much deeper and 
has been digitized,
enabling a detaield comparison with the radio and other images.

\section{Results for W~44\label{w44sec}}

\begin{figure}
\figcaption{
{\bf FIGURE NOT INCLUDED IN TEX FILE. SEE FILE 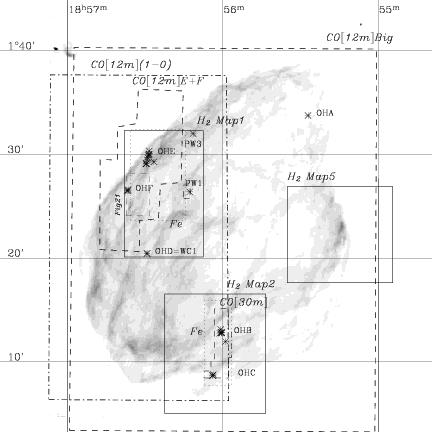.}
Finder chart for W~44 showing the locations of the three Palomar
images of H$_2$ emission (Map1, Map2, and Map5) as solid rectangles,
the OH masers from \citet{claussen} and `Prominent Wing' (PW)
and `Wing Candidate' (WC) positions from \citet{setathesis} as asterisks.
The locations of mm-wave CO observations are shown as dashed rectangles.
The smaller, dotted boxes within Map1 and Map2 (and
all of Map5) were observed in the \ion{Fe}{2} filter.
These symbols and rectangles are superposed on the
radio image from \citet{W44radio} and a J2000 coordinate grid.
\label{w44finder}
}
\end{figure}

To get oriented within W~44, Figure~\ref{w44finder} shows the radio image
together with outlines of the observed regions. 
The radio emission arises from a limb-brightened
shell that breaks into prominent filaments, many of which
appear to emanate from the southeast. 
The radio emission
is generally much brighter on the eastern hemisphere, but
there is a knot of bright radio emission on the westernmost
portion of the shell as well. 

\subsection{Millimeter-wave results for W~44}

\begin{figure}[th]
\epsscale{1}
\figcaption[f3.eps]{
{\bf FIGURE NOT INCLUDED IN TEX FILE. SEE FILE 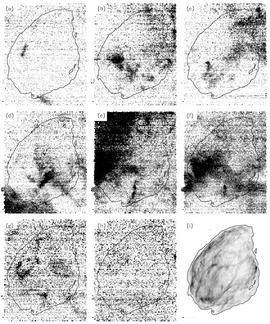.}
Images of W44 in the \cotwo\ line. Each panel
shows the integrated emission over a different velocity range, corresponding
to each major component of the CO line in the surveyed region:
{\it (a)} 94.9--97.5 km s$^{-1}$,
{\it (b)} 80.6--88.4 km s$^{-1}$,
{\it (c)} 72.8--79.3 km s$^{-1}$,
{\it (d)} 54.6--59.8 km s$^{-1}$,
{\it (e)} 45.4--52.0 km s$^{-1}$,
{\it (f)} 38.9--42.8 km s$^{-1}$,
{\it (g)} 29.8--37.6 km s$^{-1}$, 
{\it (h)} 15.5--25.9 km s$^{-1}$.
Panel {\it (i)} is the radio image, in the exact same projection
and scale.
\label{ccsum}}
\end{figure}

The \cotwo\ osbervations with the 12-m telescope covered the
entire remnant (with low sensitivity).
Figure~\ref{ccsum} shows intensity maps in different velocity
ranges. Each velocity range corresponds (in central velocity and width)
to one of the several spectral components that are present in this field.
The giant molecular cloud associated with W~44 appears
in Figures~\ref{ccsum}{\it{e--f}}. The cloud clearly extends past the
edges of our image, in particular to the north and east, consistent with
the giant molecular cloud in Figure~\ref{comini}.
The spectrum averaged over the 
portion of the giant molecular cloud in our image 
reveals a simple, Gaussian line
profile centered at 46.6 km~s$^{-1}$, with a linewidth of $4.1\pm 0.2$
km~s$^{-1}$, a peak brightness temperature of 6.9 K, and a line
integral W[\cotwo] of 30 K~km~s$^{-1}$. 
Using $N($H$_2)/W[$\coone$]=3\times 10^{20}$ cm$^{-2}$~K$^{-1}$~km$^{-1}$s
\citep{coh2ref}, this cloud has a column density 
$N($H$_2)=1.1\times 10^{22}$ cm$^{-2}$.
The column density inferred from absorption of
X-rays from the interior of the remnant is $N($H$)=2\times 10^{22}$ cm$^{-2}$
\citep{rho44};
assuming most of the gas is molecular, this would correspond to 
$N($H$_2)=1\times 10^{22}$ cm$^{-2}$ and agrees well with the
column density inferred from the CO observations, suggesting that
much of the parent cloud is in front of the remnant.
The velocity of this cloud is consistent with galactic 
rotation at the longitude and distance of W~44.
This is the ambient molecular cloud with which W~44 is interacting and
from which the progenitor star probably formed.

The northeastern ridge of bright radio emission is closely
parallel to the surface of the molecular cloud as
traced in Figure~\ref{ccsum}{\it{e}}.
It appears
that the W~44 progenitor exploded within, and just to the 
west of the center of, its parent molecular cloud. 
There is ambient gas from the parent all around the remnant. 
Figure~\ref{ccsum}{\it{f}} shows a piece of the
west of the remnant; the interaction of the remnant with this gas 
probably explains the very bright, western radio knot, for which we
present new observations of shocked gas below.
The presence of ambient molecular gas all around W~44 makes it a
special case of a supernova within a molecular cloud.
A similar situation was found in 3C~391, though its
progenitor was relatively closer to the edge of its parent
molecular cloud or even just outside \citep{wilner}.
Before the supernova, the  progenitor of W~44
would have generated only a very small \ion{H}{2} region, even if
it had been an O star, because the
surrounding medium is relatively dense. 

In addition to the ambient cloud, the CO observations reveal several
other features that can be distinguished in the maps at
velocities different from the ambient cloud.
Figure~\ref{ccsum}{\it{a}} shows two very distinct
clouds at 95 km~s$^{-1}$. The spectra of these clouds shows they have
very narrow linewidths, comparable to our resolution. The lines appear
in both \coone\ and \cotwo. 
The northeastern 95 km~s$^{-1}$ clump is located within the radio contours
of the remnant, while the southern one is just south of the remnant.
Using the rotation curve of \citet{burton_rotcurve}, 
the highest allowed velocity at this galactic longitude is 
95 km~s$^{-1}$, meaning they would have to be near the tangent point
(5--7 kpc ) if their velocity were only due to galactic rotation.
The narrow \coone\ linewidths ($<1.5$ km~s$^{-1}$) indicate that the 
clumps are cold, and probably not from shocked gas.
It is possible that these clouds are interacting with the remnant, but
based on the available evidence, we suspect that they are background
clouds unrelated to W~44.

Molecular gas currently undergoing shocks should have a linewidth greater
than that of the parent molecular cloud.
Such broad molecular line regions appear bright in at least 4 panels
of Figure~\ref{ccsum}.
Figure~\ref{ccsum}{\it{g}} shows emission at velocities less positive
than the ambient gas, due to gas that is approaching the Sun faster
than the ambient cloud.
Three features can be seen in this image: a set of peaks near and just
inside the eastern radio rim, a peak in the south-central part
of the remnant, and a large faint region inside the western part of the 
remnant. The latter region appears only in panel {\it{g}} and is apparently
a narrow CO line, due to cold gas and not necessarily associated with 
the remnant. 
The northeastern and southern regions, however, appear in multiple
panels. 
The spectra of these regions show broad molecular lines extending mostly
toward lower velocities, meaning the shocks are moving toward us. 
The line of sight from the Sun therefore passes
through the ambient gas, the shock front, the cooling gas, and then
the remnant interior. 
The high X-ray absorption, consistent with the total parent
cloud column density, supports the
idea that the bulk of the ambient molecular gas is located in
between the remnant and the Sun.
It appears the progenitor exploded on the
far side and just west of a dense portion of the parent cloud.

\begin{table}
\caption[]{Positions for Maps made at IRAM 30-m$^a$}\label{irampostab} 
\begin{flushleft} 
\begin{tabular}{rccrrl} 
\hline
Position & RA (2000) & Dec & $\Delta$ RA & $\Delta$ Dec & Description \\ \hline \hline
W 44:OHE & 18 56 28.1 & +01 29 59 & 0    & 0    & OH maser \citep{claussen} \\
     OHF & 18 56 36.7 & +01 26 32 & 129  & -207 & OH maser \citep{claussen} \\
     PW1 & 18 56 12.7 & +01 26 25 & -230 & -214 & Prominent CO(1-0) wing \citep{setathesis}; `Wing 1' \citep{seta04}\\
     PW3 & 18 56 11.7 & +01 27 55 & -246 & 124  & " \\
W 44:OHA & 18 52 55.2 & +01 29 51 & 0    & 0    & OH maser \citep{claussen} \\
W 44:OHB & 18 56 01.2 & +01 12 47 & 0    & 0    & "\\
     OHC & 18 56 03.7 & +01 08 44 & 37   & -243 & " \\
W 44:OHD & 18 56 29.4 & +01 20 26 & 0    & 0    & " \\
\hline
\end{tabular} 
$^a$ Reference position: offset (412$''$, -3621$''$) from W~44:OHE
\end{flushleft} 
\end{table}  

The IRAM 30-m observations reveal the shocked gas at higher
angular resolution and sensitivity.
The regions observed with the IRAM 30-m were selected using
the radio image, 
OH 1720 MHz maser positions \citep{claussen}, and 
Nobeyama 45-m observations \citep{seta04} as a guide.
Table~\ref{irampostab} shows the coordinates around which small
maps were made and the locations of
some other points of interest within these maps.

\begin{figure}[th]
\figcaption[f4.eps]{
{\bf FIGURE NOT INCLUDED IN TEX FILE. SEE FILE 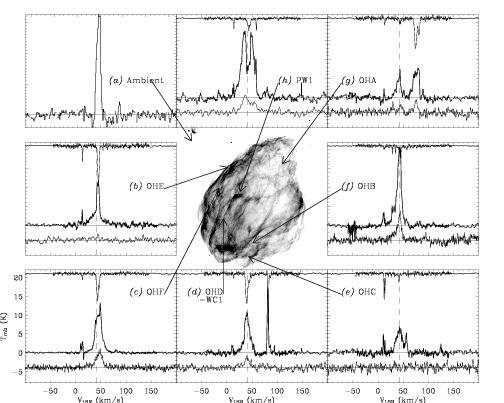.}
Spectra of \cotwo, \cstwo, and \cothirteen\ lines 
toward 8 positions in W~44, with their locations indicated on
the superposed radio continuum image (as in Fig.~\ref{w44finder}). 
Panel {\it (a)} shows the spectrum of the ambient gas outside the remnant.
In the other panels, the
thick spectrum with a baseline at 0 K is \cotwo, the thin spectrum
with a baseline at -4 K is the \cstwo\ line (with intensity multiplied 
by 10), and the thin line with a baseline at +21 K is the \cothirteen\ line 
(with intensity multiplied by -3).
\label{manyfig}}
\end{figure}

Figure~\ref{manyfig} shows the spectra obtained toward 6 OH maser locations
within W~44. Toward all positions, there is a broad
molecular line. The \csfive\ emission is only detected from a limited number
of positions, and the \csthree\ spectra are mostly too noisy to
be useful. The broad emission lines are mixed with narrower components,
sometimes at the same velocity and sometimes shifted. The presence of
the broad emission line confirms that at least part of the emission
from W~44 is due to shocked gas. Earlier observations by \citet{setaw44}
also showed broad emission lines in \coone, and we confirm that the
positions they identified as having prominent wings do indeed
have broad \cotwo\ lines. Using the higher rotational transition
has some advantages, making our data somewhat cleaner than the earlier
observations. Specifically, the upper energy level is relatively more
excited in the shocked gas than in the unshocked gas, increasing the
brightness of the shocked gas relative to the unshocked gas. Further,
the line-of-sight absorption is stronger for the 1-0 transition,
which can be absorbed by cold gas with substantial populations in
the ground state, than for the 2-1 transition. Neither of these effects
are dramatic improvements, because the $J=2$ level can be excited
by the unshocked gas (leading to narrow emission lines from the cold
gas along the line of sight) and the $J=1$ level has a substantial
population in the unshocked gas (leading to narrow absorption lines
from the cold gas along the line of sight). However, the combined
effects, together with the higher angular resolution at the higher
observing frequency, make these new data significantly cleaner
in tracing the shocked gas.

\begin{figure}[th]
\epsscale{0.5}
\figcaption[f5.eps]{
{\bf FIGURE NOT INCLUDED IN TEX FILE. SEE FILE 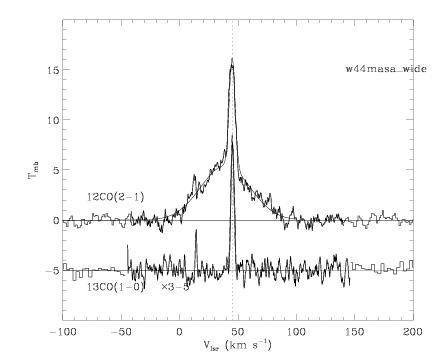.}
\cotwo\ and \cothirteen\ spectra toward
W~44:OHE. The narrow component in the \cotwo\ spectrum 
aligns exactly with the peak in the \cothirteen\ spectrum, 
suggesting that the cold gas does not absorb the emission from
the shocked gas (which makes the wide component); the cold 
gas is probably behind the shocked gas.
\label{w44masa_wide_1213}}
\epsscale{1}
\end{figure}

To investigate the distribution of the shocked gas relative to the unshocked
gas, we made small maps around a few positions that showed broad
emission lines.
Figure~\ref{w44masa_wide_1213} shows the \cotwo\ spectrum for a position
near W~44:OHE with a clear, broad component and a superposed narrow emission
component. The full-width-at-half-maximum (FWHM) of the narrow component
is 5.2 km~s$^{-1}$, typical of the ambient molecular cloud around W~44.
The FWHM of the broad component is 30.5 km~s$^{-1}$, which is far wider
than ambient gas and is due to shocks with speeds equal to or 
greater than (due to projection) 30.5 km~s$^{-1}$
into dense gas.
The amplitude of the broad component near W~44:OHE varies
significantly from position to
position, while the narrow component is widespread.
The location where the broad component is brightest is not coincident
with an OH maser. A similar result was found in the broad molecular line
region in 3C~391, 
where an OH maser located between the peaks of the broad and narrow 
components \citep{rr02}.

Another position of interest is W~44:PW1, 
the location of a `prominent wing,' also called `Wing 1,'
which was found from Nobeyama 45-m \coone\ observations
\citep{setathesis,seta04}.
Figure~\ref{manyfig}{\it (h)} shows the \cotwo\ spectrum and 
the $^{13}$CO(1-0) spectrum toward this position. The \cotwo\ emission
is clearly broad, but with a double-humped structure that could 
be mistaken for multiple, moderately-broad components along the
line of sight. The double-humped structure is actually
due to a deep absorption trough cut into a single broad emission
component. The $^{13}$CO(1-0) emission, which traces the
total column density and is dominated by the cold gas along the
line of sight, matches very well the center and width of the
apparent `notch' cut out of the \cotwo\ spectrum. The optical depth
of the absorbing CO in the J=1 level is $\tau >1$ to produce this trough.
Such an optical depth is consistent with the column density
of the parent molecular cloud (and its location in front of
the remnant) as discussed above.
Inspecting the spectra near W~44:PW1, the narrow absorption component 
transforms into a narrow emission component in locations
where there is no broad molecular line emission. 
This can be explained by a
 combination of geometry and physical conditions. 
When there is bright emission from hot, shocked gas behind cold,
unshocked gas, an absorption component appears. Then, when the
column density of shocked gas is small enough that its brightness
temperature is lower than that of the shocked gas, or the cold gas
is behind the hot gas, the narrow component appears in emission.

\begin{figure}[th]
\epsscale{1}
\figcaption[f6.eps]{
{\bf FIGURE NOT INCLUDED IN TEX FILE. SEE FILE 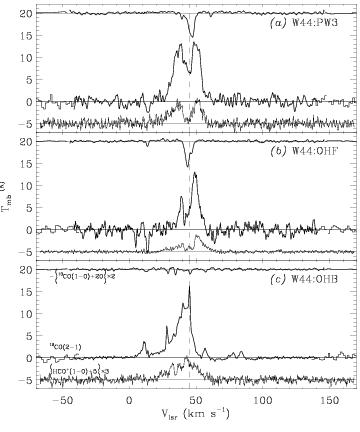.}
$^{12}$CO(2-1), HCO$^+$(1-0), and 
$^{13}$CO(1-0) spectra toward three positions in W~44.
In each panel, the lower spectrum is HCO$^+$(1-0), 
multiplied by a factor of three and shifted downward by 5 K;
the middle spectrum is $^{12}$CO(2-1); and the upper spectrum
is $^{13}$CO(1-0), inverted and multiplied by a factor of two
and shifted upward by 20 K.
In panels {\it (a)} and {\it (b)}, the
deep troughs in the $^{12}$CO(2-1) and HCO$^+$(1-0) spectra
near the parent cloud velocity (indicated by the dashed line)
align with the peaks in $^{13}$CO(1-0) spectrum, suggesting that
cold gas in front of the shocked region absorbs the hot, shocked gas. 
\label{hcop_new}}
\end{figure}

The depth, velocity, and location of the molecular absorption features 
suggests they are due to pre-shock gas.
Figure~\ref{hcop_new} shows high-quality spectra
of \cotwo, \hcop, and \cothirteen.
The HCO$^+$ spectra are similar to the \cotwo\ spectra, with deep 
absorption near the line center and even broader wings. The HCO$^+$ 
line is even more susceptible
to absorption than \cotwo\ because it is a ground-state transition
with a larger dipole moment. The optical depth in the core of
the \hcop\ line is $\sim 2$, with a width $\sim 10$ km~s$^{-1}$
consisting of multiple components. The required column density of 
HCO$^+$ in 
the ground state is $N[$HCO$^+]=1.03\times 10^{12} \tau\Delta v$ cm$^{-2}$
which amounts to $2\times 10^{13}$ cm$^{-2}$. 
If the abundance of HCO$^+$ relative to H$_2$ is 
$2\times 10^{-9}$ typical of molecular clouds \citep{liszt},
then the inferred column density
in the absorbing cloud is $N[$H$_2]\sim1\times 10^{22}$ cm$^{-2}$.
This column density agrees very well with that 
of the parent molecular cloud, based on the \coone\ brightness
and X-ray absorption.
Because it is at the same velocity and spatial location 
as the remnant, they must be close together,
with the absorbing gas being the unshocked portion of the parent
molecular cloud along the line of sight.

\begin{figure}
\vspace{6truein}
\figcaption[f7.eps]{
{\bf FIGURE NOT INCLUDED IN TEX FILE. SEE FILE 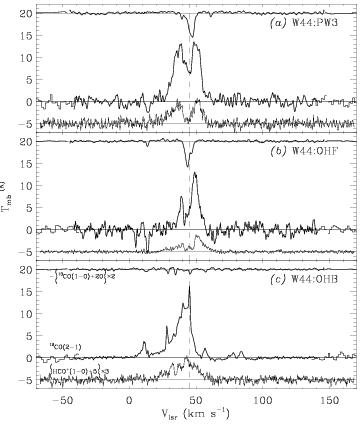.}
Grid of spectra near W~44:OHB.
The grid is labeled with the offsets (in arcsec) East and North of the nominal
position W~44:OHB. There are both wide and narrow components throughout this
region. The bottom of the figure covers the edge of the remnant, and the bottom
right is somewhat outside the remnant, giving an idea of the baseline,
random emission around the remnant. Around (0$''$,80$''$), the emission appears to 
be two medium-wide components, but this is probably due to absorption by
cold gas in front of the remnant. The range of appearances of these spectra 
is due to a complicated mix of emission and absorption with different angular
distributions and velocity widths. There are two OH masers in this region,
at (0,0) and (37,-243).
\label{w44map2co21map}}
\end{figure}

The CO emission in the southern portion of W~44 was observed in some detail.
It appears as a broad-line region in the NRAO 12-m observations:
in Figure~\ref{ccsum}{\it e} it appears as a roughly north-south
filament starting just inside the southernmost boundary of the
remnant. 
The region  contains the maser complexes W~44:OHB and W~44:OHC.
The area mapped with the IRAM 30-m is
indicated in Figure~\ref{w44finder} as
a dashed box labeled `CO [30-m]'.
Figure~\ref{w44map2co21map} shows the grid of spectra,
which vary dramatically from
position to position. A broad emission line is clearly evident 
in a region extending from offsets (0,0) to (80,-260). 
In the spectra at the top of Figure~\ref{w44map2co21map}, the
emission arises from a set of narrow components. It is 
possible that this is actually broad-line emission being absorbed
by foreground gas, as was found for W~44:PW1, W~44:PW3, and
W~44:OHF; but we only have the $^{13}$CO or HCO$^+$ observations that
disentangle the absorption and emission for a few lines of sight
in the remnant. The angular resolution of the 30-m telescope
in the \cotwo\ line is about $10.5''$, much smaller than the
region over which broad emission lines are observed,
so it is clear that the emitting regions are extended.
The profiles tend to have a wider wing on the low-velocity side,
suggesting that the gas is being accelerated toward us.

\subsection{Near-infrared results for W~44}

\begin{figure}
\figcaption{
{\bf FIGURE NOT INCLUDED IN TEX FILE. SEE FILE 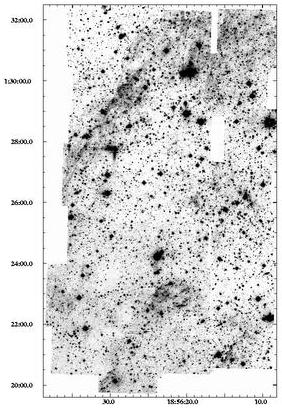.}
H$_2$ 2.12 $\mu$m image of the northeastern portion of W~44 (Map1).
The portion of the supernova remnant covered by this image
is indicated in Fig.~\ref{w44finder}. 
The greyscale ranges from 0 (white) to 
$1\times 10^{-4}$ erg s$^{-1}$ cm$^{-2}$ sr$^{-1}$ (black) in this and 
the other near-infrared images in this paper.
\label{w44map1}}
\end{figure}

Turning to the near-infrared observations,
the 2.12 $\mu$m H$_2$ images are striking in appearance, with an 
intricate network
of filaments and some isolated bright clumps. 
The three regions that were observed in the near-infrared also 
contain regions with
bright \ion{O}{1} 63 $\mu$m emission \citep{rr96}, OH masers \citep{claussen},
and broad molecular lines \citep{seta04}. 
In the W~44 northeastern image (Fig.~\ref{w44map1}),
a ridge of H$_2$ emission follows the edge of the radio supernova remnant,
which runs diagonally across the upper left portion of the image.
The H$_2$ emission contains bright filaments and arcs as well as a
more diffuse component that generally envelops the brighter
features. Some of the filaments are extremely narrow and probably
represent individual shock fronts seen close to edge on;
these are discussed in \S\ref{edgeonshock} below.
The other structures, and the diffuse emission, are probably an amalgam
of multiple shock fronts seen from a range of angles.
Significant shocked H$_2$ emission is detected interior to the radio shell, 
for example in the lower-central portion of the image.

\begin{figure}
\figcaption{
{\bf FIGURE NOT INCLUDED IN TEX FILE. SEE FILE 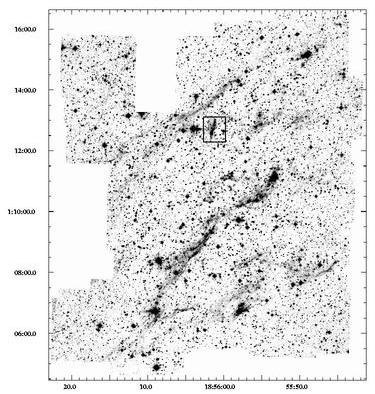.}
H$_2$ 2.12 $\mu$m image of the southern portion of W~44 (Map2).
The portion of the supernova remnant covered by this image
is indicated in Fig.~\ref{w44finder}. 
The clump of exceptionally bright H$_2$ emission discussed in the text
is indicated by a black rectangle.
\label{w44map2}}
\end{figure}

The W~44 southern image (Fig.~\ref{w44map2} greyscale, 
Fig.~\ref{w44map2color} color) is the most 
detailed, having 4 times as much integration per pixel as the
other H$_2$ images. Two bright sets of filaments run diagonally across 
the field, with fainter filaments (all in the same general direction) 
distributed throughout the field. 
A very bright knot of H$_2$ emission lies
just below the upper filament in Fig.~\ref{w44map2}. This is not
a star or an artifact; it is just burned out in the greyscale image.
There are some narrow filaments that appear to be individual edge-on
shock fronts, but for the most part the bright filaments have 
significant substructure indicating they comprise multiple shock
fronts.

The W~44 western field (Fig.~\ref{w44map5}) contains some very bright
filaments, again roughly parallel to the edge of the remnant, which
runs roughly north-south at this location. It is possible to see
individual shock fronts in this image; counting across the
image along a single line (from the upper left to the lower right),
there are at least 15 individual shock fronts. This shows
clearly that the pre-shock medium was highly structured with
gaps between regions containing the gas with density appropriate
to generate bright H$_2$ after being shocked.

\begin{figure}
\figcaption{Color version of the H$_2$ 2.12 $\mu$m image of the southern 
portion of W~44 (Map2).  {\bf FIGURE NOT INCLUDED IN ASTRO-PH SUBMISSION. SEE ApJ ARTICLE}
\label{w44map2color}}
\end{figure}

\begin{figure}
\figcaption{
{\bf FIGURE NOT INCLUDED IN TEX FILE. SEE FILE 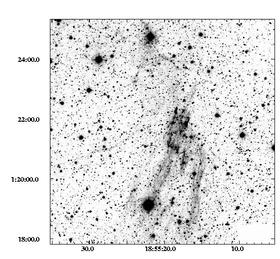.}
H$_2$ 2.12 $\mu$m image of the western portion of W~44 (Map5).
\label{w44map5}}
\end{figure}

\clearpage

\section{Results for W~28\label{w28sec}}

\begin{figure}
\figcaption{
{\bf FIGURE NOT INCLUDED IN TEX FILE. SEE FILE 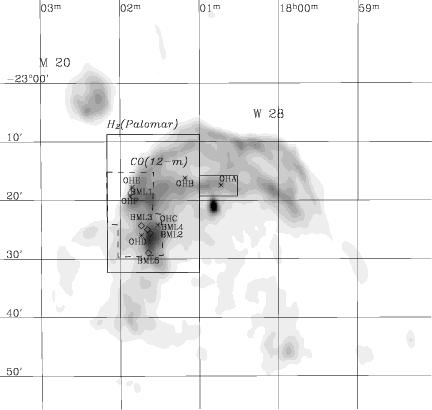.}
Finder chart for W~28 showing the locations of the Palomar
images of H$_2$ emission as a solid polygon,
and the locations of CO observations with the 12-m as a dashed polygon,
the OH masers from \citet{claussen} as asterisks,
and the BML positions as diamonds,
superposed on the
radio image from \citet{W28radiodubner}.
\label{w28finder}}
\end{figure}

An orientation chart showing the regions observed toward W~28 is
shown in Figure~\ref{w28finder}. The main geographical features
of W~28 are a semi-circular radio shell that is brightest in
the north and east, and a bar (or inner shell) that extends
east-west.
The observed regions include most of the OH masers 
\citep{claussen}, the bright radio ridges, and the \cothree\ ridge
\citep{arikawa}.

\subsection{Millimeter-wave results for W~28}

\begin{figure}[th]
\figcaption[f13.eps]{
{\bf FIGURE NOT INCLUDED IN TEX FILE. SEE FILE 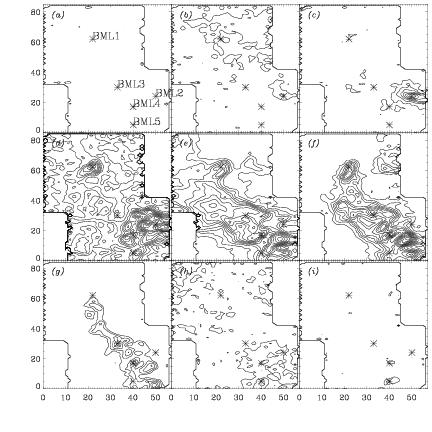.}
\cotwo\ 
maps of the eastern ridge of W~28 integrated over 9 different velocity ranges:
{\it (a)}      48.28 to      63.88,
{\it (b)}	   40.48 to	  46.98,
{\it (c)}	   28.77 to	  37.88,
{\it (d)}	   15.12 to	  28.12,
{\it (e)}	   4.074 to	  7.975,
{\it (f)}	  -4.376 to	  2.124,
{\it (g)}	  -17.37 to	 -5.027,
{\it (h)}	  -33.63 to	 -18.02,
{\it (i)}	  -53.13 to	 -34.28 km~s$^{-1}$.
Each of these velocity ranges corresponds to a peak in a spectrum, somewhere
within the mapped region.
The locations of 5 broad-molecular-line (BML) regions are indicated
by asterisks and are labeled in panel {\it (a)}.
Contours are at integer multiples of 14 K~km~s$^{-1}$.
\label{w28cm}}
\end{figure}

\begin{figure}[th]
\figcaption[f14.eps]{
{\bf FIGURE NOT INCLUDED IN TEX FILE. SEE FILE 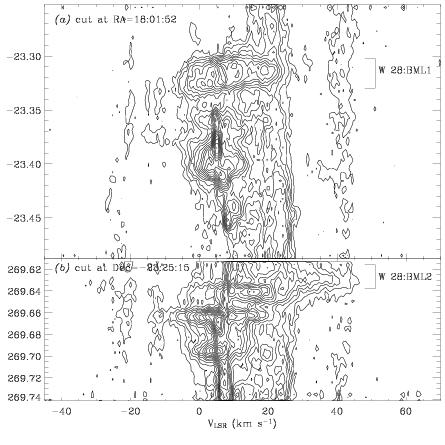.}
\cotwo\ position-velocity maps: {\it (a)} cut at
constant RA $18^h01^m52^s$; {\it (b)} cut at Dec 
$-23^\circ 25^\prime 15^{\prime\prime}$.
The locations of two broad-molecular line regions that are intersected by the cuts
are indicated. 
Contours are at integer multiples of 2 K. The positions are labeled in
decimal degrees, both for declination {\it (a)} and
right ascension{\it (b)}.
\label{w28pv}}
\end{figure}

The 12-m CO observations covered 
a $9^\prime\times 13^\prime$ region centered on 
the bright eastern radio ridge,
containing many OH 1720 MHz masers \citep{frail94}.
Figure~\ref{w28cm} shows a set of velocity-integrated images. 
Broad molecular line (BML) regions are those that appear in multiple images,
in particular those away from the line core.
Five broad-molecular-line regions can be identified in these images;
Table~\ref{w28bmltab} lists their positions and gaussian
fits to the line profiles.
The BML regions are even more easily seen in the position-velocity images;
Figure~\ref{w28pv} shows position-velocity slices through the spectral data.
The exceptionally wide velocity dispersion of the BML regions compared to 
the ambient
gas (which has typical width less than 7 km~s$^{-1}$ FWHM) is evident.
It is also evident that the BML regions are well-defined peaks. There
is more extensive, broad-line-emitting gas, but the BML regions
in Figures~\ref{w28cm} and~\ref{w28pv} are clumps (probably containing
significant structure unresolved in our $30''$ beam).
A position-velocity slice through the northern radio shell
in \cothree\ was shown by \citet{arikawa}
[their Fig. 2], and the \cothree\ image [their Fig. 3]
shows extended BML emission, so the peaks listed in our
Table~\ref{w28bmltab} are only a subset of the BML regions
in W~28.

\begin{figure}
\figcaption{
{\bf FIGURE NOT INCLUDED IN TEX FILE. SEE FILE 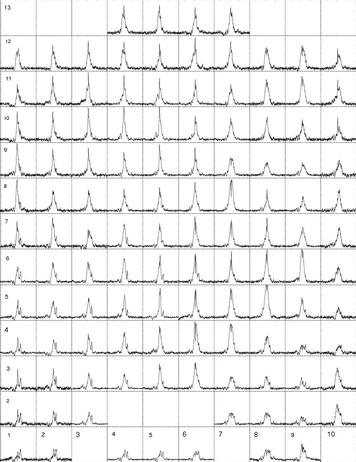.}
Grid of \cotwo\ spectra toward broad-molecular-line regions
in W~28, with a spacing of $20''$. Within this
$10\times 13$ grid, W~28:BML4 is in 
Column 7, Row 8 (counting from the lower left).
W~28:BML3 is in the upper left
of this grid. OH 1720 MHz masers are at Maser (Column, Row):
OHC2 (10,12), OHD2 (5,5), OHD3 (5,4) \citep{claussen}.
\label{w28panels}}
\end{figure}

The IRAM 30-m observations of W~28 reveal the bright CO 
emission at higher angular resolution. 
Figure~\ref{w28panels} shows a grid of spectra centered on some of
the brightest CO emission, including the broad-line positions 
BML3 and BML4.
The broad-line emission is widespread, and it is deeply cut
by narrow absorption at approximately the same central velocity
(8 km~s$^{-1}$).
The absorption is particularly deep toward W~28:BML4, with an
optical depth $> 1$ and width $\sim 1$ km~s$^{-1}$ FWHM.
The absorption is patchy, but cold gas at this velocity is present throughout
the region; emission lines from the cold gas 
are seen in the spectra with weaker broad-line emission.
Three OH masers are included within the boundaries of this map;
there is nothing noticeably special about the spectra centered on 
the OH masers.

\begin{figure}
\figcaption{
{\bf FIGURE NOT INCLUDED IN TEX FILE. SEE FILE 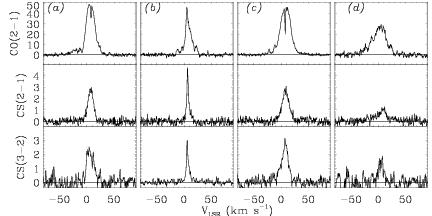.}
Spectra of {\it (a)} W~28:BML2, 
{\it (b)} W~28:BML3, {\it (c)} W~28:BML4, and {\it (d)} W~28:BML5 in the
\cotwo\ {\it (top)}, \cstwo\ {\it (middle)}, and \csthree\ {\it (bottom)} lines.
\label{w28many}}
\end{figure}

Figure~\ref{w28many} shows the spectra of 4 of the 5 broad-molecular-line 
(BML) regions from Table~\ref{w28bmltab} in the \cotwo, \cstwo, 
and \csthree\ lines.
The \cotwo\ lines are characterized by a moderately broad 
(20--30 km~s${-1}$)
component with narrow absorptions biting notches out of the broad
line profile for W~28:BML2 and BML4. Table~\ref{w28bmltab} shows
the brightnesses and velocities of the various lines, determined 
using gaussian fits. For BML2 and BML4, the \cotwo\ absorption notches
are listed separately as negative gaussians. For BML3, the line profiles
were best fit with a sum of broad and narrow gaussian; this does not
mean that there are two separate components, as the combination of
foreground absorption and non-gaussian profile can give the
impression of two components even when there is only a single
emission component. Recall that both W~28 and W~44 are located far
from the Sun and near the galactic plane, so there are several
clouds between the remnants and the Sun.
For BML5, the \csthree\ line is weak and the wings are probably lost in 
the  noise.

\begin{table}
\caption[]{Broad Molecular Line regions in W~28}\label{w28bmltab} 
\begin{flushleft} 
\begin{tabular}{rlcccl} 
\hline
RA  & Dec & $\int T_{mb} dv$ & $\langle V \rangle$ & $\Delta V$\\
\multicolumn{2}{c}{(2000)} & (K~km~s$^{-1}$) & (km~s$^{-1}$) & (km~s$^{-1}$)  & line \\ \hline \hline
W~28:BML1 &  & & & & \\
18:01:52.0&-23:18:47  &     466  &     11.7  &     21.3 & \cotwo \\
          & 	    &     19   &      4.3  &     18.2 & \cstwo \\ 
          & 	    &     33   &      3.6  &     28.8 & \hcop \\ 
          & 	    &     3.2  &      9.6  &     21.9 & SiO($2\rightarrow 1$) \\ 
W~28:BML2  && & & & \\ 
18:01:37.4&-23:25:44  &     970  &	 7.1  &	18.9 & \cotwo \\
          & 	    &     -21  &     8.0   &    1.5 & \cotwo\ absorption\\
          & 	    &    36   &	  6.6    & 12.1	& \cstwo \\
	    & 	    &    37   &	  6.6    & 15.6	& \csthree \\
W~28:BML3  && & & & \\
18:01:44.1&-23:24:24  &	  422  &	  9.1  &	 17.1 & \cotwo\ broad \\
          & 	    &     100  &      5.8  &     3.8 & \cotwo\ narrow \\
          & 	    &     14  &	   7.5  &	  8.9 & \cstwo\ broad \\
          & 	    &     8   &	   6.1  &	  2.4 & \cstwo\ narrow \\
          & 	    &     14  &	   7.5  &	  8.9 & \csthree\ broad \\
          & 	    &     8   &	   6.1  &	  2.4 & \csthree\ narrow \\
W~28:BML4  && & & & \\ 
18:01:39.5&-23:25:04  &   1090  &	 7.6  &	24.1 & \cotwo \\
          & 	    &     -29  &     7.6   &    1.2 & \cotwo\ absorption\\
          & 	    &     43   &     7.6 &  15.6 & \cstwo \\
          & 	    &     41   &     7.3 &  15.0 & \csthree \\
W~28:BML5  & & & & & \\ 
18:01:38.8&-23:29:03  &  840  &	 2.4  &   30.4 & \cotwo \\
          & 	    &      25  &    5.5   & 29.2   & \cstwo \\
	    & 	    &      17  &    5.4   & 11.1:  & \csthree \\
\hline
\end{tabular} 
\end{flushleft} 
\end{table}

The mm-wave line ratios for the W~28:BML1--BML5 are generally similar
to each other: \cstwo/\cotwo$\sim 0.04$, \csthree/\cstwo$\sim 1.0$. 
In 3C~391:BML1, the CS lines weaker, relative to than \cotwo, by a
factor of 2, but the ratio \csthree/\cstwo\ is similar, suggesting
similar excitation conditions of $T_k\sim 100$ K and 
$n({\rm H}_2)\sim 10^5$ cm$^{-3}$ \citep{rr96}. The abundances
of CS and HCO$^+$ are similar to those found in 3C~391 and IC~443
\citep{ziurys} to within a factor of 3.

\begin{figure}
\figcaption{
{\bf FIGURE NOT INCLUDED IN TEX FILE. SEE FILE 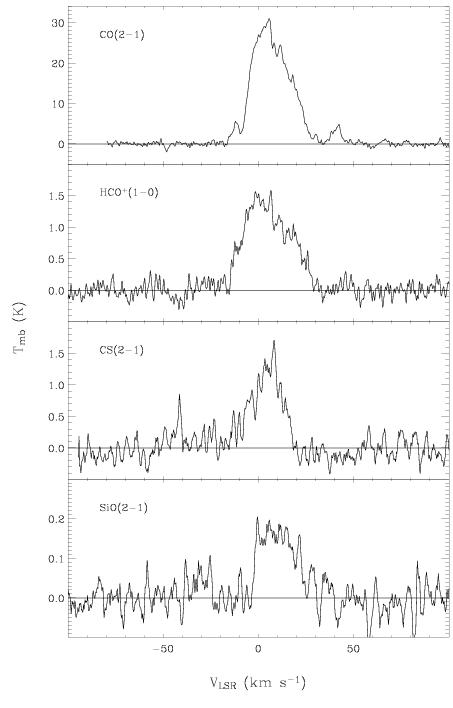.}
Spectra of SiO emission from W~28:BML1, taken with the IRAM 30-m.
This spectrum combines 30 minutes of on-source integration. The spectra
of 3 other molecules toward the same position (and smoothed to the same
angular resolution) are shown for comparison.
\label{w28sio}}
\end{figure}

The SiO($2\rightarrow 1$) line was detected toward W~28:BML1:
Figure~\ref{w28sio} shows the spectrum and
Table~\ref{w28bmltab} shows the line fit.
While the emission is weak, the SiO line profile is
similar to other molecular transitions observed toward the same 
position.
Of the other positions observed in this
line, neither W~44:OHB nor W~44:OHF were detected, with $T_A^*<0.12$ K.
Because Si is locked in grains in quiescent
interstellar gas, the presence of Si in the gas phase indicates either
grain destruction or advanced grain-surface chemistry.
\citet{ziurys} detected SiO toward IC 443 and argued that the gas responsible
for it has strongly enhanced SiO abundance, $100\times$ the upper limits
of SiO abundance in dark clouds. \citet{shilke} modeled the production of
SiO in shocks and showed that the abundance is strongly enhanced by
Si liberated from the shocked grain mantles and reacting with other molecules
in the post-shock gas. To get the SiO line brightness we observed requires
that the shocks are faster than 20 km~s$^{-1}$, consistent with the
observed line width of 21 km~s$^{-1}$ (representing a lower limit to the shock velocity).
Gas-phase Si was also detected in the far-infrared spectra of both W~28 and
W~44, through the 34.8 $\mu$m fine-structure line of Si$^+$ \citep{rr00}. 


\subsection{Near-infrared results for W~28}

\begin{figure}
\figcaption{
{\bf FIGURE NOT INCLUDED IN TEX FILE. SEE FILE 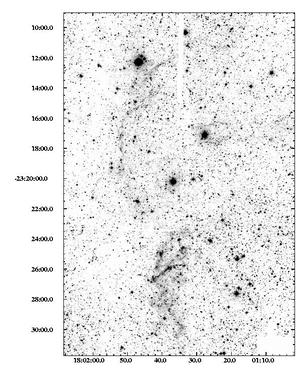.}
H$_2$ 2.12 $\mu$m image of the eastern portion of W~28.
\label{w28h2}}
\end{figure}

Let us now inspect the near-infrared image.
 The eastern portion of W~28 (Fig.~\ref{w28h2}) 
contains the brightest H$_2$ emission we observed
in either remnant.
There are large-scale H$_2$ arc complexes, 
one in the southeast and one in the northeast.
The southeast large-scale arc is a very bright spur running
from the easternmost portion of the arc toward the northwest.
The bright bars contain the broad-molecular-line regions
W~28:BML2 (brighter bar running diagonally SE/NW) and
W~28:BML4 (upper bar running more N/S);
these bright, broad, linear
features have an average surface brightness of
$10^{-3}$ erg~cm$^{-2}$~s$^{-1}$~sr$^{-1}$.

Fainter H$_2$ arcs are distributed throughout the region,
generally along and parallel to the radio ridge, which comprises a
north-south bar that turns roughly at a right angle into an
east-west part at the location of the northeastern H$_2$ arc.
The northeastern and southeastern H$_2$ arcs are connected
by fainter emission; as a whole this H$_2$ ridge runs almost exactly 
parallel to the ridge of CO wing emission in Figure~\ref{w28cm}{\it (g)}. 
The CO wing region W~28:BML1 is located just west of a relatively
bright H$_2$ filament in the northeastern arc; this filament
is probably due to a single, bright, edge-on shock propagating into a
dense molecular core.

\clearpage

The northern radio ridge, which runs east-west across 
the region covered by Figure~\ref{w28h2}, about 1/3 of the
way from the top,
contains the ridge of
shocked \cothree\ emission that was detected by \citet{arikawa}.
There are some thin H$_2$ filaments that probably represent edge-on
shocks. A pair of bright, nearly parallel filaments 
bracket a bright star near the junction between the north-south
and east-west radio ridges. There are at least 3 shock fronts at
this individual location. 

\clearpage

\section{Discussion}

\subsection{Correlation between the shocked CO and H$_2$\label{seccoh2}}

\begin{figure}
\figcaption{
{\bf FIGURE NOT INCLUDED IN TEX FILE. SEE FILE 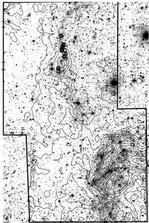.}
Overlay of CO emission (thin contours) on the
H$_2$ image (greyscale) of the eastern portion of W~28. The region observed
in CO is bounded by the thick bounding box. 
Small squares indicate the locations of OH 1720 MHz masers.
\label{w28h2co}}
\end{figure}

The clumpy H$_2$ emission is spatially
associated with the broad-line CO emission. 
Examples of this are in the southern part of W~44,
where the bright elliptical region of H$_2$ emission 
(Fig.~\ref{w44map2}) corresponds to the broad-line
CO-emitting region near W~44:OH~B (Fig.~\ref{w44map2co21map}).
Also the bright H$_2$ emission in W~28 exists
in the region of broad CO lines; Figure~\ref{w28h2co} shows the 
CO overlaid on the H$_2$ image.
The correspondence between H$_2$ and CO is not
perfect, but the overall location of the CO and H$_2$ are similar
and some features agree in detail. In particular,
the bright CO bars in the southern portion of the image
are both associated with H$_2$. Individual H$_2$ filaments are narrow
($\sim 1$ to $2''$), so they are highly diluted in the CO beam ($30''$)
and may be detected with better angular resolution or sensitivity.

A good correspondence between H$_2$ and broad CO was seen
for IC~443 \citep{ic443h2,rho443}.
Such correspondence was also seen in 3C~391,
where the H$_2$ broke into clumps that
are likely pre-existing
dense cores in the parent molecular cloud \citep{rr00}. 
Such condensations
are expected in star-forming clouds such as the parent clouds of
W~28 and W~44. Since the progenitors of W~28 and W~44 were
massive stars, forming rapidly and existing as stars briefly,
it is like that there is continuing formation of 
lower-mass stars from the same clouds,
in accordance with the initial mass function \citep{imf}.
The blast waves from the W~28 and W~44 progenitors are
propagating into these environments, which span a range of densities.
The volume density inferred from the CO excitation discussed above,
and from  the detection of far-infrared H$_2$O, OH, and
high-$J$ CO lines \citep{rr98}, is $n_C > 10^5$ cm$^{3}$.
Pre-stellar cores have significant regions with density $\sim 10^5$ cm$^{-3}$
on a timescale $\sim 10^{5}$ yr \citep{andrecore}. 
Shocks into these regions will be non-dissociative, and the H$_2$
emission arises from collisionally-heated molecules behind the shock
\citep{DRD}.

\subsection{Physical properties of the shocked CO}

\begin{figure}[th]
\epsscale{.5}
\epsscale{1}
\figcaption[f20.eps]{
{\bf FIGURE NOT INCLUDED IN TEX FILE. SEE FILE 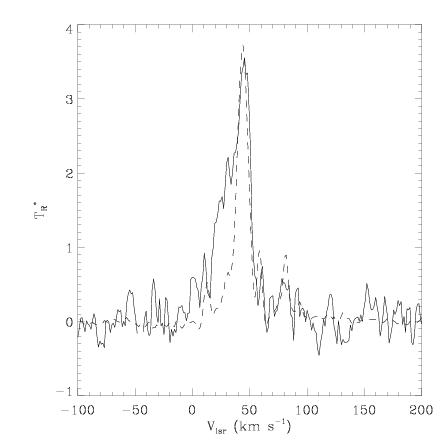.}
Spectra of \cotwo\ [solid line] and \coone\
[dashed line, scaled by a factor of 0.5] averaged over the portions of 
the W~44 (0.4 arcmin$^2$) with the most prominent broad molecular 
line wings. It is clear that the line wings are much more prominent in 
the $2\rightarrow 1$ line, while the narrow component is essentially identical in 
the two lines.
\label{co21_10}}
\end{figure}

The CO observations clearly differentiate between the shocked and
unshocked gas. In addition to the broad line profile, the brightness
ratios among the various spectral lines is different in the shocked
gas as compared to the ambient gas. 
This is because the shocked gas is
warmer (and often denser) than the unshocked gas, so that the relative 
brightness
of shocked gas is higher in \cotwo\ than in \coone. 
Figure~\ref{co21_10} illustrates this clearly. The narrow component
is present in both spectral lines, with identical
width. The brightness ratio 
$R_{21}\equiv W[CO(2-1)]/W[CO(1-0)] = 0.7\pm0.1$,
using the line integral of gaussian fits to the narrow component.
The line ratio for the ambient gas is typical for molecular clouds
in the inner galaxy, where for example a value of $0.74\pm 0.02$ was
observed for clouds in the Galactic center region, and
0.77 and 0.66 were observed
for two giant molecular clouds in Orion,
consistent with cold, optically thick gas,
for example $T\sim 10$ and $n\sim 10^3$ cm$^{-3}$ 
\citep{oka,sakamoto94}.

The broad 
component is much fainter in \coone\ than in \cotwo:
$R_{21}=3.5\pm 0.9$, using the line integral over -20 to +65 km~s$^{-1}$ 
after subtracting the narrow component.
The line ratio $R_{21}$ for the shocked gas is much higher than that of
ambient molecular gas, due to a combination of heating and compression.
A similarly high $R_{21}$ was seen for the shocked gas in 
HB~21 \citep{koohb21} and IC~443 \citep{setaw44}.
That the line ratios are different is not surprising, if
the broad line arises from molecules surviving a C-type shock 
while the narrow line is mostly ambient, unshocked gas.
The utility of \cotwo/\coone\ as a tracer of shocked gas has also 
been discussed by \citet{setaw44}.
The actual line ratio could be even
higher in the shocked gas, if there is significant structure 
on angular scales smaller than the 
beam---which is likely, given the fine-scale structure of the 
shocked H$_2$ images---so it is 
clear that there is a dramatic difference between the properties of
the shocked and ambient gas.

We can constrain the physical properties of the broad-line
emitting gas using $R_{21}$ and the isotopic brightness ratio
$R_{13/12}\equiv W[^{13}CO(1\rightarrow 0)]/W[^{12}CO(1\rightarrow 0)]$.
From Figure~\ref{w44masa_wide_1213}, 
the upper limit to the brightness of $^{13}$\coone is 2\% 
of the brightness of \cotwo,
which converts, using $R_{21}=3.5$, to a brightness
ratio $R_{13/12}< 0.07$. Assuming an isotopic
ratio of $^{13}$CO/$^{12}$CO=50 \citep{co1312_ratio}, the upper
limit to the optical depth is $\tau[$\coone$]<3$.
Using the optically thin limit, and balancing collisions and radiative 
transitions for the lowest several levels of CO,
we find that the temperature of the
shocked CO $\sim 100$ K, independent of density,
and a lower limit to the density is $n($H$_2) > 4\times 10^3$ cm$^{-3}$.

An independent estimate of the properties can be obtained from
the CS observations.
For lines of sight with detected \csthree\ and \cstwo, we measure
$R_{32}^{CS}\equiv W[CS(3\rightarrow 2)]/W[CS(2\rightarrow 1)]\simeq 1$
(Table~\ref{w28bmltab}), which
requires high volume densities,
$n($H$_2)>9\times 10^4$ cm$^{-3}$ for $T\simlt 100$ K.

The CO and CS line ratios can be
explained by gas with $n($H$_2)=2\times 10^5$ cm$^{-3}$ and 
$T=100$ K, which are the conditions we inferred
from a multi-level analysis of CS lines from the shocked gas in
in 3C~391, where the \csfive\ line was also observed \citep{rr99}. 
It is clear that the broad emission
lines are due to warm gas, heated by the shock. This is important
because fast shocks, with $v_s>40$ km~s$^{-1}$ dissociate H$_2$
molecules \citep{drainemckee}. The width of the broad molecular
lines, $\sim 20$ km~s$^{-1}$, is probably 
indicative of the velocity of the shocks it experiencing.
Warm, broad-line
gas is most easily explained by non-dissociative
shocks \citep{DRD}.

\subsection{Relationship of OH masers to shocked CO and H$_2$}

Many shock-excited OH 1720 MHz masers are located within W~44 and W~28.
All of the OH masers are located in or near H$_2$ emission, and all of the masers are
in or near broad CO emission. This clearly supports the idea that
OH masers are signposts of interaction between supernova remnants and
molecular clouds \citep{mitchell}.
But the H$_2$ and CO images are far from a one-to-one
correspondence with OH masers: 
many locations of bright H$_2$ have no OH masers, and the
CO and H$_2$ peaks only rarely coincide with masers. 


Figure~\ref{w28h2co} compares the locations of
33 OH masers \citep{claussen} in W~28 with the H$_2$ and CO images.
The OH masers closely trace one of the filaments of H$_2$ emission,
in the northeast. This is not the brightest H$_2$ filament, nor is it
obviously special in any other way. But the conditions within this 
filament---its viewing geometry, or its location with respect to the 
bright radio synchrotron emission that the maser can amplify---are evidently 
much better for creating maser emission
that other, comparable portions of the remnant.
Examples of regions where OH masers lie along H$_2$ filaments are seen in
the northeast portion and souther portion (Fig.~\ref{w44h2rad}) of W~44 as well. 
The spatial alignments are too
detailed to be coincidence. Counterexamples of similar filaments with no
OH masers, or OH masers not located on bright H$_2$ filaments, 
are present throughout both remnants. 
Wherever there are OH masers, though, there is H$_2$ emission.
Thus it appears that OH masers are a 
very non-linear tracer of shocked molecular gas. The presence of OH masers
clearly indicates molecular shocks, but a lack of OH masers is not informative.


\subsection{Shock thickness\label{edgeonshock}}

The H$_2$ images of W~28 and W~44 reveal bright, complex regions
as well as a network of fainter, narrow filaments. 
Figure~\ref{w44map1fil} shows an example of a region with 
many narrow filaments.
The individual shock fronts in W~28 and W~44 are barely resolved in
the near-infrared H$_2$ images, with angular widths $\simlt 0.7$--$4''$. 
If they are interpreted as edge-on shock fronts, then we are
observing shock fronts with thicknesses of 3--$12\times 10^{16}$ cm. 
This is really an upper limit, because the narrowest shocks are
unresolved at the limit of atmospheric seeing, and all of these shocks
have some inclination with respect to the line of sight. In what 
follows, we compare the shock thickness to theoretical models
for non-dissociative and dissociative shocks.

The observed widths agree with the predictions for noon-dissociative
shocks with properties as were invoked above
to explain other properties of the gas (CO and CS line ratios).
From equation 3.12 of \citet{drainemckee}, using a preshock magnetic field
strength of $b n^{1/2}$ $\mu$G where $n$ is the preshock gas
density in cm$^{-3}$ and $b$ is a dimensionless field strength
of order unity, the shock thickness 
\[
L\sim 1.3\times 10^{16} b (10^{-4}/x) (10^2/n) \,\,{\rm  cm},
\]
where $x=n_i/n$ and $n_i$ is the ion density.
For preshock gas densities between $n=10^2$ and $10^4$ cm$^{-3}$
and corresponding
ionization fractions between $x=10^{-4}$ and $10^{-7}$, respectively,
the shock thickness is in the range 1--10 $\times 10^{16}$ cm.
\citet{DRD}, in Figs. 1--2 of their paper, predict the structures of
of 25 km~s$^{-1}$ shocks into gas with pre-shock H-nucleon 
density $10^2$ and $10^4$ cm$^{-3}$.
The width of the region
where H$_2$ would be heated, but not converted into more
complex molecules like H$_2$O, is smaller than the total shock
thickness estimated by the equations above: reading from
their figures, the H$_2$ emission regions 
are $\sim 1 \times 10^{16}$ cm.
The observed widths of the shock fronts are consistent with 
non-dissociative shocks into gas with densities of order $10^4$ cm$^{-3}$.
Shocks into gas with pre-shock density $10^6$ cm$^{-3}$
(if they are steady shocks, which may
not occur on the timescales of the remnants we are observing) 
are predicted to be wider than observed, and Figure 3 
of \citet{DRD} shows most
of the cooling will come from coolants other than H$_2$.

\begin{figure}
\figcaption{
{\bf FIGURE NOT INCLUDED IN ASTRO-PH SUBMISSION. SEE ApJ ARTICLE}
Blow-up of a portion of the northeastern rim of W~44 in 
the 2.12 $\mu$m H$_2$ line. The region depicted occupies the
lower portion of Fig.~\ref{w44map1}; it is outlined as a dashed
rectangle, labeled {\it Fig21}, in Fig.~\ref{w44finder}.
\label{w44map1fil}}
\end{figure}

\def\ergcmssr{erg~s$^{-1}$~cm$^{-2}$~sr$^{-1}$}

Faster, dissociative shocks, can also produce H$_2$ emission 
from molecules reforming behind the shock.
\citet{HM89} calculated the width of the H$_2$ reformation region 
to be (their equation 3.4) 
\[
5\times 10^{17} b (100/n) [ 1 + 100 e^{-(6\times 10^6/n)^{0.2}} ]\,\,{\rm cm},
\]
for a shock speed of 100 km~s$^{-1}$, as was inferred from the width 
of the \ion{O}{1} 63 $\mu$m emission line \citep{rr00}.
The observed shock thickness
can be produced if the density $n\sim 10^3$ cm$^{-3}$. 
Thus, based only on the observed thickness, there are two types of
shock that can produce structures with the thickness observed:
non-dissociative shocks into gas with density $\simgt 10^4$ cm$^{-3}$
and dissociative shocks into gas with density $\sim 10^3$ cm$^{-3}$.

\subsection{Relation between molecular and ionic shocks}

\begin{figure}
\figcaption{\
{\bf FIGURE NOT INCLUDED IN TEX FILE. SEE FILE 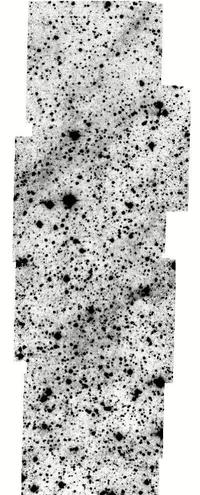.}
ion{Fe}{2} 1.644 $\mu$m image of the southern portion of W~44.
This is a subfield of the region covered by the H$_2$ image in 
Figure~\ref{w44map2}; the location in the remnant is indicated in 
Figure~\ref{w44finder}.
\label{w44map2fe}}
\end{figure}

Portions of the remnants were observed in the \ion{Fe}{2} 1.644 $\mu$m
filter. While the H$_2$ emission is bright and extended, with
intricate filaments as well as bright blobs, the \ion{Fe}{2} emission
is weak and relatively diffuse. 
Figure~\ref{w44map2fe} shows the \ion{Fe}{2} image of the southern
portion of W~44, 
revealing two diagonal filaments,
qualitatively similar to the H$_2$ image (Fig.~\ref{w44map2}) but
with the \ion{Fe}{2} filaments much more diffuse and shifted north of the 
H$_2$ filaments. The separation of the \ion{Fe}{2} and H$_2$ emission
is not particularly surprising, considering that the physical conditions
of the shocks traced by these lines is very different.
However, the rough alignment of the filaments and the fact that they 
are offset in the direction toward the center of the remnant is
very suggestive. We might expect that a fast shock hitting the
surface of a moderately-dense molecular cloud would produce \ion{Fe}{2}, 
from the destroyed grains,
as was seen in IC~443 \citep{rho443} and 3C~391 \citep{rr02}.
Then it is conceivable that the H$_2$ would be reformed molecules
behind the shock, as was predicted theoretically by \citet{HM89}
and tentatively observed in one remnant by \citet{koow51}.
However, the H$_2$ emission in this image of W~44 is located
further from the remnant center than the \ion{Fe}{2} emitting region,
whereas reformed molecules should be behind the shock.
The observed separation between the molecular and ionic
shocks is $2\times 10^{18}$ cm, which is much larger than the expected size 
of the molecular reformation region behind a fast shock:
for $n_0=10^3$ cm$^{-3}$ and $v_s=100$ km~s$^{-1}$,
$z_{1/2}\simeq 10^{16}$ cm \citep{HM89}.
The \ion{Fe}{2} traces a combination of grain destruction 
(to get Fe in the gas phase) and hot and dense regions (to excite
the upper energy levels). 
It is unlikely that the \ion{Fe}{2} emitting
region could end up further behind the shock than the H$_2$ emission.

Therefore, we interpret the \ion{Fe}{2} and H$_2$ filaments
as tracing independent shocks: the \ion{Fe}{2} traces fast, grain-destroying
shocks into $n< 10^3$ cm$^{-3}$ gas, while the H$_2$ traces denser
shocks. The reason for their rough correspondence and occasional
parallel alignment on the sky is probably
reflective of the pre-shock structure of the molecular cloud.
When approaching a dense portion of the molecular cloud,
the blast wave first encounters the less-dense material at its surface 
yielding the ionic shocks. Slower shocks then 
begin propagating into the denser material, yielding the molecular shocks. 

\subsection{Shock brightness}

The H$_2$ (1$\rightarrow 0$) S(1) line brightness of the individual, 
filamentary shock fronts in Figure~\ref{w44map1fil}
is about $5\times 10^{-5}$ \ergcmssr. Brighter features, such as
the bright clump in Figure~\ref{w44map2}, range up to 
$2\times 10^{-4}$ \ergcmssr. Let us compare these values to
those predicted by shock models consistent with other observed
properties. The predicted face-on brightness of nondissociative
shock fronts into $n=10^4$ cm$^{-3}$ and $V_s=25$ km~s$^{-1}$ is
$\sim 3\times 10^{-5}$ \ergcmssr\ \citep{DRD}, 
and the predicted face-on brightness
of dissociative, J-shocks into $n=10^3$ cm$^{-3}$ and $V_s=100$ km~s$^{-1}$ is
$\sim 2\times 10^{-5}$ \ergcmssr\ \citep{HM89}.
Both of these predictions are comparable to, but a factor of
2--3 fainter than, the brightness of the 
individual, filamentary shock fronts. 
The theoretical models are for face-on shocks, which are fainter
than edge-on shocks. It is quite plausible for there to be 
a factor of 2--3 geometric correction between face-on 
(theoretical) and more nearly edge-on (observed) geometries.
It is a rather remarkable coincidence
that both the thickness and brightness of the near-infrared
H$_2$ emission, from two very different types of shocks, is
so similar.

There is a significant
difference in the physical properties of H$_2$ behind the different
shocks. Behind a non-dissociative, C-type shock, the H$_2$ molecules
are heated without being dissociated; there must exist a critical
type of shock that can just barely dissociate H$_2$ that will
yield the highest possible excitation. The neutral gas temperature
behind such shocks reaches $10^3$ K then cools rapidly,
leading to bright emission from the S(3) and S(5) pure 
rotational lines \citep{DRD}. 
In contrast, behind a faster shock, the H$_2$ molecules are 
destroyed and the H ionized; the molecules reform $\sim 10^{17}$ cm
behind the shock where the temperature is $10^{2.5}$ K and lower.
The cooling is somewhat slower because of the lower density,
leading to bright emission from from the S(7) and S(9) pure 
rotational lines \citep{HM89}. 
Let us now take models consistent with the observed
near-infrared 1--0 S(1) line brightness and 
consider the {\it Infrared Space Observatory} data on the
pure rotational lines.
For one line of sight each in remnant, near W~28:OHF and W~44:OHE, 
we detected the S(3) and S(9) lines in the $14''\times 20''$ 
spectrometer aperture \citep{rr00}.
The ratios S(9)/S(3) are 0.1 and 0.5 for W~28:OHF and W~44:OHE, 
respectively, with $<10$\% uncertainties.
The predicted line ratios are S(9)/S(3)=10 for the
dissociative shock and S(9)/S(3)=0.02 for the 
non-dissociative shock. The observed line ratio is
inconsistent with {\it both} shock models, so we infer
that there must be different types of shock producing
different lines. Using the theoretically predicted 
ratios of the S(3) and S(9) lines, the S(3) line arises
almost exclusively from the non-dissociative shock ($> 94$ \%),
while the S(9) line arises largely from the dissociative
shock ($> 83$ \%). Thus both types of 
shock are expected to contribute to the near-infrared H$_2$ emission 
with comparable brightness averaged over large beams.

To separate between the dissociative and non-dissociative shocks, 
we must turn to evidence other than the shock width and brightness.
The filamentary H$_2$ emission is unlikely to arise from
very dense cores ($n_0>10^4$ cm$^{-3}$), because the filaments are
coherent and long ($> 2$ pc). If the filamentary H$_2$ emission
is from a shock front propagating into a coherent three-dimensional
region with this size and density, the pre-shock gas would have
a mass $\gg 10^4 M_\odot$ and would have
been easily visible in the millimeter-wave lines that trace
such gas. In fact, the \cstwo\ spectra do not show such large-scale
dense material, revealing instead only relatively isolated cores
with much smaller size. 
The distribution of pre-shock molecular gas is complicated,
and the new images of shock fronts propagating into giant
molecular clouds presented in this paper deserve more 
complicated theoretical models to elucidate both the
shock physics and pre-shock gas distribution.

\subsection{Molecular shocks and cosmic rays}

Supernova remnants are the most plausible source for galactic cosmic
rays \citep{drainemckee,jonesellison}. Most attention has been focused
on young supernova remnants, where the shocks are strong enough to 
accelerate thermal electrons to relativistic energies. 
W~28 and W~44 are very bright radio
continuum sources \citep{clarkcaswell}; they are the 6$^{th}$ and 7$^{th}$
brightest supernova remnants out of 232 remnants in the
Green catalog.
The synchrotron radiation could be bright for several reasons:
the strength of the magnetic field in the dense material with which the 
blast waves are interacting, local 
acceleration of particles to high energies, re-acceleration
of existing high-energy particles, or some combination of these
effects. Particle acceleration (or reaccelration) will depend
on the physical conditions of the pre-shock gas and the strength
of the shock. 

If all the present-day shocks into different regions are being driven by 
the same ram pressure, then we can discern which
types of pre-shock gas are the most likely origin for particle
acceleration or re-acceleration by comparing images that
trace the various types of shock to the radio image.
For young remnants, the radio and X-ray images sometimes show
very close correspondence---for example the images of Tycho compared
by \citet{blandford}, or the near-infrared synchrotron, radio,
and X-ray images of Cas A compared by \citet{rhocasa}---so 
the cosmic ray acceleration is associated with the main
blast wave. 
But for mixed-morphology remnants like W~28 and W~44, whose defining
characteristic is the contrasting X-ray and radio morphology, we
must look to shocks that do {\it not} emit copious X-rays to find
the source of cosmic ray acceleration.

\begin{figure}
\figcaption{
{\bf FIGURE NOT INCLUDED IN TEX FILE. SEE FILE 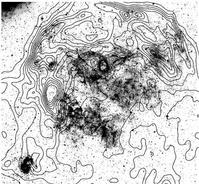.}
Overlay of radio contours onto the H$\alpha$ image (greyscale)
for W~28. The very bright black spot in the upper left corner
is the Trifid Nebula. The radio contours surround the H$\alpha$ emission
from W~28, which is the box-shaped region in the center of the image.
The H$\alpha$
image suffers from some extinction. In particular, there is
a thick band running diagonally NE-SW that appears in extinction
both in the H$\alpha$ and red continuum images.
The radio contours range from 0.1 to 1 mJy/beam in 12 equal steps.
\label{w28harad}}
\end{figure}

First, let us compare the H$\alpha$ and radio images.
We assume that the near-infrared H$_2$ emission and broad CO emission
trace shocks into dense gas ($> 10^3$ cm$^{-3}$), and  the
H$\alpha$ image traces shocks into lower-density gas ($n_A\sim 5$ cm$^{-3}$).
Figure~\ref{w28harad} shows that the radio image is quite
different from H$\alpha$ for W~28, with the radio shell surrounding the
centrally-filled H$\alpha$ emission. 
This morphology was already noted by \citet{vandenbergh}.
Some faint H$\alpha$ emission is associated with the outer
radio shell, for example along the northernmost radio arc
and the northwestern portion of the remnant \citep{W28radiodubner}. 
But the vast majority of the
H$\alpha$ emission arises from the interior of the remnant,
where the radio emission is very weak. In particular, the
brightest radio emission arises from a very bright bar that runs
north-south and forms the eastern boundary of the remnant. 
The H$\alpha$ emission fills the interior of the remnant and
ends abruptly at the radio bar. The second-brightest radio 
feature is a bar that runs east-west in the
northern part of the remnant. The bright H$\alpha$ emission
in the interior again terminates at this radio bar as well.
The third-brightest radio feature is a thin arc of fainter
emission that forms the far northern boundary of the remnant;
this is where some faint H$\alpha$ emission appears associated 
with the radio emission.
Thus it appears that, for the most part, the inner boundary
of the radio shell delineates
the outer boundary of the H$\alpha$ emitting region. 
The morphology of the
radio and H$\alpha$ emitting regions is also different: while the radio
can be described as a set of large, coherent arcs running
along the edge of the remnant, the H$\alpha$ emission is highly
structured with many very thin filaments, some of which 
run orthogonal to the radio arcs. The difference in fine structure
is difficult to assess with the present data, however, because their
angular resolution of the radio data is much lower than that of the
H$\alpha$.
Similar results obtain for W~44, though the
H$\alpha$ emission is weaker. \citet{rho44} showed
that the H$\alpha$ emission is interior to the radio shell
and has different morphology. \citet{giacani} showed that
there are some portions of the W~44 where the H$\alpha$ and
[\ion{S}{2}] appear to follow each other and the radio, though there
is also optical emission from regions with no corresponding
radio emission. The situation for both W~28 and W~44 is
that some regions have corresponding H$\alpha$
and radio emission, but for the most part (and especially
in the interior of the remnant), the
radio and optical images are very different.

\begin{figure}
\figcaption{
{\bf FIGURE NOT INCLUDED IN TEX FILE. SEE FILE 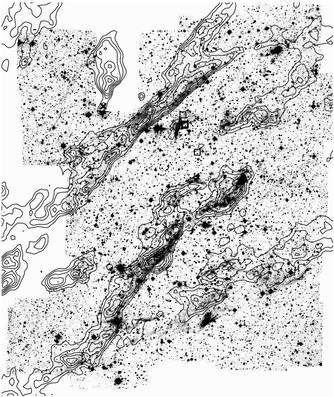.}
Overlay of radio contours onto the H$_2$ image (greyscale)
for a $10.1'\times 11.7'$ in the southern portion of W~44;
(the same region in Figs.~\ref{w44map2} and~\ref{w44map2color}).
Contours are evenly spaced from 0.3 to 2.5 mJy/beam.
The radio image was spatially filtered by subtracting a version
smoothed to $4'$, which removes the diffuse emission 
(which has a brightness of 1.5 mJy/beam over most of the field).
The radio beam size is $15''$. Squares denote the positions
of OH 1720 MHz masers, and the `{\bf $\times$}' indicates the
position of the pulsar B1853+01.
\label{w44h2rad}}
\end{figure}

Next, let us compare the H$_2$ and radio images. Figure~\ref{w44h2rad}
shows radio contours overlaid on the H$_2$ image of the southern 
portion of W~44. Both images are dominated by two diagonal 
filaments running SE-NW across the field. The orientation
of the filaments matches remarkably well in these two images
of disparate emission mechanisms. The radio emission from the
southernmost of the two bright filaments overlays the H$_2$ filament
very closely. Even the relatively faint radio and H$_2$ features
have very good correspondence, throughout this image. 

The brightest patch of H$_2$ emission, which is just south
of the northern filament in Figure~\ref{w44h2rad} and contains the
W~44:OHB maser complex, has a faint 
radio counterpart of about 1.5 mJy/beam brightness.
The radio/H$_2$ ratio is at 3 times less in the H$_2$
clump than in the H$_2$ filaments.
The shocks into the highest-density gas,
traced by the `clumpy' type of H$_2$ emission (as opposed to the long,
thin filaments), OH masers, and wide CO lines, has apparently
a lower synchrotron emissivity than the filaments.

For W~28, a similar correspondence is found between H$_2$ and
radio emission, though the angular resolution of the radio 
image is much worse. The bright H$_2$ emission in the SE
portion of Figure~\ref{w28h2} is precisely where the radio
emission in the remnant is brightest in its north-south bar.
The shocked CO emission was shown to be well correlated with
the radio image by \citet{arikawa} and \citep{W28radiodubner}.

Because the radio and shocked molecular filaments are well 
correlated, we suggest that the bright radio emission from 
W~28 and W~44 is due to 
interaction between the remnant and molecular gas.
Such an association is somewhat surprising, because it would 
seem that the fastest shocks would be most closely associated
with energetic particle acceleration. But strong magnetic fields
in dense gas, and the presence of stronger shocks into the lower
density gas, combine to provide the particle 
acceleration and synchrotron enhancement, respectively.
\citet{bykov} has addressed this issue theoretically.
The present-day shocks in W~28 and W~44 may not be strong enough to
be primary accelerators of cosmic rays, though
fast shock fronts are capable of accelerating thermal electrons
to relativistic energies \citep{jonesellison}. 
What we observe as enhanced radio emission is
relativistic particles that emit bright synchrotron radiation 
as they spiral around the
strong magnetic field in the compressed regions behind
the molecular shocks. This mechanism was proposed by 
\citet{blandford}.
For a remnant of given explosion energy and present size, 
the radio surface brightness depends primarily on the filling
factor, $f$, of dense material behind the shock:
$\Sigma \propto B^{-0.3} f$,
where $B$ is the magnetic field in the dense gas.
The radio map is therefore predicted to trace the dense gas,
as is indeed observed in Figure~\ref{w44h2rad}. 
Quantitatively, the radio brightness of W~44 ranges from 
3--10 MJy~sr$^{-1}$ (where 1 MJy = $10^{-17}$ erg~s$^{-1}$
cm$^{-2}$~Hz$^{-1}$) at 1420 MHz.
The observed brightness is somewhat higher than the model predictions,
even for a high filling factor of dense gas; this is not
surprising because the model assumes a low filling factor
and neglects the effect of the high density on the remnant
evolution, which becomes important at higher filling factors.
Thus our new observations appear to confirm the concept
proposed by \citet{blandford}. 

Based on far-infrared and millimeter-wave spectroscopy, we have
suggested there are three types of gas that produce the main
observable features of supernova remnants interacting with
molecular clouds \citep{rr00}.
Table~\ref{tab:cartoon} describes the physical
properties of these three types of gas.
The three types, $i$, of gas with density $n_i$, shock velocity $V_i$,
and filling factor $f_i$, will contribute to
the synchrotron emission proportional to $n_i^{1.75} V_i f_i$,
assuming magnetic field $B_i \propto n_i^{0.5}$. 
The kinetic energy in their shocks is proportional to 
$p_{ram,i} f_i\propto n_i V_i^2 f_i$. 
The filling factors in Table~\ref{tab:cartoon} 
are very rough approximations based on the fraction
of remnant surface that is traced by each phase.
The ram pressure of shocks into the different phases is similar
but not equal: the observed properties of the
denser shocks seem to require higher ram pressure \citep{rr96}.
This effect may have been theoretically explained by \citet{chevalier}.
The radio emission arises from the intermediate-density molecular phase, 
in agreement
with our finding that the H$_2$ and radio images are associated.
Phase $M$ dominates the mass, although phase $A$ contains the 
most kinetic energy.

\begin{deluxetable}{lcccl}
\tablewidth{0pt}
\tabletypesize{\scriptsize}
\tablecaption{Three types of shocks in molecular clouds}\label{tab:cartoon}
\tablehead{
&\colhead{Atomic} &\colhead{Molecular} &\colhead{Clump} &
}
\startdata
\cutinhead{A. Properties of the pre-shock gas:}
description     & atomic, inter-clump & molecular           & clump, core    & \\
tracer          & H~I                 & low-$J$ CO          & CS, protostars & \\
density, $n_0$  & 5                   & 200                 & $2\times 10^4$ & (cm$^{-3}$) \\
fill factor, $f$ & 0.9                & 0.1                 & $10^{-4}$      & \\
mass            & 800                 & 5000                & 300            & ($M_\odot$) \\
\cutinhead{B. Observed properties of the shocked gas:}
description     & shell               & filaments           & clumps & \\
tracer          & radio, X-ray        & radio, [\ion{O}{1}], H$_2$ & H$_2$, broad molecular lines & \\
velocity, $V_S$ & 500                 & 100                 & 25            & (km~s$^{-1}$) \\
ram pressure, $p_{ram}$ & 3           & 5                   & 20     & ($10^{-8}$ dyne~cm$^{-2}$) \\
energy, $E_{kin}$ & 4                 & 1                   & 0.005  & ($10^{51}$ erg) \\
density          & 20                 & $3\times 10^3$      & $2\times 10^5$ & (cm$^{-3}$) \\
synchrotron\tablenotemark{a}      & 0.01               & 0.7                 & 0.3    & \\
\cutinhead{C. Theoretical expectations for the shocked gas:}
shock type       & very fast          & dissociative        & magnetohydrodynamic & \\
main coolant     & metastable lines   & fine structure lines, grains & molecular lines & \\
$I[$H$_2$1-0 S(1)$]$ & ...            & $3\times 10^{-5}$   & $2\times 10^{-5}$ & \ergcmssr \\
H$_2$ S(3)/1-0 S(1) & ...             & 0.5                 & 150 & \\
H$_2$ S(9)/1-0 S(1) & ...             & 4                   & 2 & \\
$I[$\ion{O}{1},63$\mu$m$]$ & $1\times 10^{-4}$ & $1\times 10^{-3}$ & $1\times 10^{-5}$ & \ergcmssr \\
\enddata
\tablenotetext{a}{Fraction of synchrotron emission produced be each type of gas.}
\end{deluxetable}

There is another possible source of energetic particles,
at least for W~44: the pulsar PSR 1853+01.
The pulsar is located just north of
Figure~\ref{w44h2rad},
and the distance from the pulsar to the northern radio filament is 
only $\sim 10^{18}$ cm. The radio filaments in W~44 seem to 
converge in the southwest portion of the remnant, and we
speculate that that energetic particles could originate from
the pulsar wind then spiral along the filaments of enhanced
magnetic field. \citet{dejager} also proposed that
the pulsar supplies the energetic particles, based on 
interpretation of the radio-to-$\gamma$-ray spectrum.

\subsection{Implications for the structure of molecular clouds}

We mentioned in the introduction that theoretical models 
\citep{cox99,chevalier} for
molecular supernova remnants explain most observable properties
as shocks into gas with a density of 5 cm$^{-3}$. 
These models are well grounded in observational data: in particular,
\ion{H}{1} 21-cm observations indicate neutral hydrogen shells
containing $\sim 10^3$ $M_\odot$ of material, suggesting
pre-shock densities $\sim 2$--5 cm$^{-3}$ for W~28 \citep{velazquez}
and W~44 \citep{koow44}.
From our millimeter-wave and infrared observations, we have
inferred shocks into gas with densities $10^3$ cm$^{-3}$ 
or higher. How can these be reconciled? 
Molecular clouds comprise regions with
a wide range of densities, from dense, star-forming cores
with $n>10^5$ cm$^{-3}$ 
to CO-emitting regions with $n> 3\times 10^2$ cm$^{-3}$,
and possibly lower if there is interclump gas.
Detailed observations of molecular clouds have shown that 
CO emission is from 
regions with volume density $n_{M}$ much higher than the 
than the path-length averaged density 
$\langle n\rangle\equiv N/L$.
For a giant molecular cloud such as the ones near
W~44 and W~28, $N\sim 2\times 10^{22}$ cm$^{-2}$ and $L\sim 100$ pc, 
so $\langle n\rangle \sim 10^2$ cm$^{-3}$.
A detailed study of the Rosette molecular cloud reveals that the
CO emission arises from clumps that occupy only 8\% of the cloud
volume; the remainder of the cloud is filled with diffuse
atomic gas with a mean density $n_{A}\sim 4$ cm$^{-3}$ \citep{williams}.
In a recent study of \ion{H}{1} absorption toward a large sample of dark
clouds, \citet{ligoldsmith} showed pervasive atomic gas within molecular 
clouds, with a volume density $n($\ion{H}{1}$)\sim 4$ cm$^{-3}$.
They interpreted the \ion{H}{1} as the product of cosmic-ray destruction
of H$_2$ in dark clouds; since the \ion{H}{1} volume density is predicted
to be independent of the H$_2$ density, the \ion{H}{1} can 
be thought of as an interclump medium in a clumpy cloud.
There is increasing evidence that the bulk of the mass of molecular clouds 
resides in clumps that fill only a small fraction of the cloud.
An IRAM key project demonstrated that CO emission arises from cells 
that do not fill the volume of the cloud, having densities $10^3$--$10^5$ cm$^{-3}$
and sizes of order 200 AU \citep{falgarone}.
Multi-level excitation studies
show that even the CO($4\rightarrow 3$) line is bright
so that the fourth rotational level is excited; the
inferred volume density $n_{M}\sim 10^{4.5}$ cm$^{-3}$
\citep{ingalls00}.


\section{Conclusions}

We have used supernova blast waves as a means of illuminating
the cloud structure. Radiative shocks into gas of different
densities have different cooling mechanisms, allowing us to separate
shocks into dense cores, moderate-density molecular gas, and 
interclump atomic gas.
Giant molecular clouds are pervaded by interclump gas
with density $\sim 5$ cm$^{-3}$, with moderate-density
CO-emitting portions occupying $\sim 10$\%, and denser
gas occupying yet smaller volume.

One of the defining characteristics of supernova remnants
in the mixed-morphology class to which W~28 and W~44 belong is their
centrally-condensed, thermal X-ray emission \citep{RP98}.
Two competing theoretical explanations for the presence of such
a large amount of interior X-ray emitting gas
involve evaporating clumps inside the remnant \citep{whitelong} and 
thermal conduction behind radiative shocks \citep{cox99}.
We now suspect that both mechanisms are operating. 
The thermal conduction model explains some interior
X-rays, assuming shocks into interclump gas with a density of 5 cm$^{-3}$.
But this model does not produce a central column density
peak with very flat temperature profile, required to match
the X-ray observations.
The combination of the radiative shock into the interclump gas
and evaporating material from the dense clumps that survive
the shock may work in combination to produce the X-ray
emitting material. 
A reservoir of dense material that can survive the initial blast wave 
is clearly present: it manifests itself through
broad molecular line regions, bright
H$_2$ clumps, and OH masers. The shocks into these clumps are much slower
than the shock into the interclump gas, which leaves
the clumps behind to evaporate in the interior. The variegated
appearance of these two supernova remnants interacting with 
molecular clouds owes to the wide range of densities already
present in the clouds, with fast shocks producing X-ray,
H$\alpha$, and \ion{Fe}{2} emission, slower shocks into
moderate-density gas producing filamentary H$_2$, and radio emission, 
and slower-yet shocks into dense cores producing CO, CS, and HCO$^+$
emission.

\acknowledgments 
This work is based on observations obtained at the Hale Telescope, 
Palomar Observatory, 
as part of a continuing collaboration between the California Institute
of Technology, NASA/JPL, and Cornell University.
We are pleased to acknowledge the assistance of Naman Bhatt, who worked
with us for a summer on this project.
This publication makes use of data products from the Two Micron All Sky Survey, which is a joint project of the University of Massachusetts and the Infrared Processing
and Analysis Center/California Institute of Technology, funded by the National Aeronautics and Space Administration and the National Science Foundation.
Observations with the 12-meter telescope were made while that telescope
was being operated by the National Radio Astronomy Observatory (NRAO).
The NRAO is a facility of the
National Science Foundation, operated under cooperative
agreement by Associated Universities, Inc.
The research described in this paper was carried out at
the California Institute of Technology under a contract with the 
National Aeronautics and Space Administration.	


\clearpage

\begin{thebibliography}{} 
 
\bibitem[Andr\'e, Ward-Thompson, \& Barsony(2000)]{andrecore} Andr\'e, P., 
Ward-Thompson, D., \& Barsony, M. 2000, in {\it Protostars and Planets IV},
eds. V. Mannings, A. P. Boss, \& S. S. Russell (Tucson: University of Arizona Press), 59
\bibitem[Arikawa et al.(1999)]{arikawa} Arikawa, Y., Tatematsu, K., Sekimoto, Y.,
\& Takahashi, T. 1999, PASJ, 51, L7
\bibitem[Bitran et al.(1997)]{bitran} Bitran et al. 1997, A \& A S, 125, 99
\bibitem[Blandford and Cowie(1982)]{blandford} Blandford, R. D., and
Cowie, L. L. 1982, \apj, 260, 625
\bibitem[Bloemen et al.(1986)]{coh2ref} Bloemen, J. B. G. M., Strong, A. W.,
Mayer-Hasselwander, H. A., Blitz, L., Cohen, R. S., Dame, T. M., Grabelsky, D. A.,
Thaddeus, P., Hermsen, W., \& Lebrun, F. 1986, \aap, 154, 25
\bibitem[Burton et al.(1988)]{ic443h2} Burton, M. G., Geballe, T. R., 
Brand, P. W. J. L., and Webster, A. S. 1988, MNRAS, 231, 617
\bibitem[Burton(1988)]{burton_rotcurve} Burton, W. B. 1988, in {\it Galactic and
Extragalactic Radio Astronomy}, eds. G. L. Verschuur and K. I. Kellermann
(New York: Springer), p. 296
\bibitem[Bykov et al.(2000)]{bykov} Bykov, A. M., Chevalier, R. A., Ellison, D. C.,
Uvarov, Yu. A. 2000, \apj, 538, 203
\bibitem[Chevalier(1999)]{chevalier} Chevalier, R. A. 1999, \apj, 511, 798
\bibitem[Clark and Caswell(1976)]{clarkcaswell} Clark, D. H., and
Caswell, J. L. 1976, MNRAS, 174, 267
\bibitem[Claussen \etal(1997)]{claussen} Claussen, M. J., Frail, D. A., Goss,
 W. M., \& Gaume, R. A. 1997, \apj, 489, 143
\bibitem[Claussen \etal(1999)]{claussenmerlin} Claussen, M. J., Goss, W. M.,
Frail, D. A., \& Desai, K. 1999, \apj, 522, 349
\bibitem[Claussen et al.(2002)]{claussen28} Claussen, M. J., Goss, W. M., 
Desai, K. M., \& Brogan, C. L. 2002, \apj, 580, 909
\bibitem[Cox et al.(1999)]{cox99} Cox, D. P., Shelton, R. L., Maciejewski, W.,
Smith, R. K., Plewa, T., Pawl, A., \& R\'ozyczka, M. 1999, \apj, 524, 179
\bibitem[Dame et al.(1986)]{dame86} Dame, T. M., Elmegreen, B. G.,
Cohen, R. S., and Thaddeus, P. 1986, \apj, 305, 892
\bibitem[Dame, Hartmann, and Thaddeus(2001)]{dht} Dame, T. M., Hartmann, D., 
and Thaddeus, P. 2001, \apj, 547, 792
\bibitem[de Jager and Mastichiadis(1997)]{dejager} de Jager, O. C.,
and Mastichiadis, A. 1997, \apj, 482, 874
\bibitem[Denoyer(1983)]{denoyer} Denoyer, L. K. 1983, \apj, 264, 141
\bibitem[Dickel, Dickel, and Crutcher(1976)]{dickel76} Dickel, J. R.,
Dickel, H. R., and Crutcher, R. M. 1976, PASP, 88, 840
\bibitem[Draine and McKee(1993)]{drainemckee} Draine, B. T., \& McKee, C. F.
1993, ARA\& A, 31, 373
\bibitem[Draine et al.(1983)]{DRD} Draine, B. T., Roberge, W. G., \& Dalgarno, A. 1983, ApJ, 264, 485 
\bibitem[Dubner et al.(2000)]{W28radiodubner} Dubner, G. M., Vel\'azquez, P. F.,
Goss, W. M., Holdaway, M. A. 2000, AJ, 120, 1933
\bibitem[Esposito et al.(1996)]{esposito} Esposito, J. A., Hunter, S. D., Kanbach, G., 
Sreekumar, P. 1996, ApJ, 461, 820
\bibitem[Falgarone et al.(1998)]{falgarone} Falgarone, E., Panis, J.-F., Heithausen, A.,
Perault, M., Stutzki, J.. Puget, J.-L., \& Bensch, F. 1998, \aap, 331, 669
\bibitem[Frail et al.(1994a)]{frail94} Frail, D. A., Goss, W. M., Slysh, V. I. 1994, ApJ, 424, L111
\bibitem[Frail et al.(1994b)]{W28radiofrail} Frail, D. A., Kassimn N. E., and
Weilner K. W. 1994, AJ, 107, 1120
\bibitem[Frail et al.(1994)]{frail44} Frail, D. A., Giacani, E. B., Goss, W. M.,
\& Dubner, G. 1994, ApJ, 464, L165
\bibitem[Frail \& Mitchell(1998)]{mitchell} Frail, D. A., \& Mitchell, G. F., ApJ, 508, 690
\bibitem[Fryer(1999)]{fryer} Fryer, C. L. 1999, ApJ, 522, 413
\bibitem[Giacani et al.(1997)]{giacani} Giacani, E. B., Dubner, G. M., Kassim, N. E.,
Frail, D. A., Goss, W. M., Winkler, P. F., and Williams, B. F. 2001, AJ, 113, 1379
\bibitem[Green(2001)]{green} Green D.A., 2001, `A Catalogue of Galactic Supernova Remnants 
(2001 December version)', 
Mullard Radio Astronomy Observatory, Cavendish Laboratory, Cambridge, United 
Kingdom (available on the World-Wide-Web at
http://www.mrao.cam.ac.uk/surveys/snrs/).
\bibitem[Hartman, R. C. et al.(1999)]{hartman} Hartman, R. C.,
Bertsch, D. L., Bloom, S. D., Chen, A. W., Deines-Jones, P.,
Esposito, J. A., Fichtel, C. E., Friedlander, D. P.,
Hunter, S. D., McDonald, L. M., and 17 coauthors 1999, ApJS, 123, 79
\bibitem[Hollenbach \& McKee(1989)]{HM89} Hollenbach, D. J., and McKee, C. F. 1989, ApJ, 342, 306
\bibitem[Ingalls et al.(2000)]{ingalls00} Ingalls, J. G., Bania, T. M., Lane, A. P., Rumitz, M.,
\& Stark, A. A. 2000, ApJ, 535, 211
\bibitem[Jones, Tielens, \& Hollenbach(1996)]{jones96} Jones, A. P., Tielens, A. G. G. M., \& Hollenbach, D. J.
 1996, \apj, 469, 740
\bibitem[Jones and Ellison(1991)]{jonesellison} Jones, F. C., and Ellison, D. C. 1991,
Space Sci. Rev., 48, 359
\bibitem[Jones et al.(1993)]{W44radio}
Jones, L. R., Smith, A., Angellini, L. 1993, MNRAS, 265, 631
\bibitem[Kaspi et al.(1993)]{kaspi28} Kaspi, V. M., Lyne, A. G., Manchester, R. N.,
Johnston, S., D'Amico, N. D., \& Shemar, S. L. 1993, ApJ, 409, L57
\bibitem[Knapp and Kerr(1974)]{knapp74} Knapp, G. R., and Kerr, F. J. 1974, 
\aap, 33, 463
\bibitem[Koo \& Heiles(1995)]{koow44} Koo, B,-C., and Heiles, C. 1995, ApJ, 442, 679
\bibitem[Koo \& Moon(1997)]{koow51} Koo, B.-C., \& Moon, D.-S. 1997, ApJ, 485, 263
\bibitem[Koo et al.(2001)]{koohb21} Koo, B.-C., Rho, J., Reach, W. T., Jung, J.,
 \& Mangum, J. G., ApJ, 552, 175
\bibitem[Li and Goldsmith(2003)]{ligoldsmith} Li, D., and Goldsmith, P. F. 2003,
ApJ, 585, 823
\bibitem[Liszt and Lucas(2000)]{liszt} Liszt, H. and Lucas, R. 2000, \aap, 355, 333
\bibitem[Lockett et al.(1998)]{lockett} Lockett, P., Gauthier, E., \& Elitzur, M.
1999, ApJ, 511, 235
\bibitem[Long et al.(1991)]{long91} Long, K. S., Blair, W. P., Matsui, Y.,
White, R. L. 1991, ApJ, 373, 567
\bibitem[MacGillivray(1998)]{macgillivray} MacGillivray, H. T. 1998, PASA, 15, 42
\bibitem[Malin(1998)]{malin} Malin, D. 1998, PASA, 15, 38
\bibitem[Oka et al.(1996)]{oka} Oka, T. Hasegawa, T., Handa, T., Hayashi, M.,
\& Sakamoto, S. 1996, ApJ, 334, 342
\bibitem[Parker \& Phillipps(1998)]{parker} Parker, Q. A., \& Phillipps, S. PASA, 15, 28

\bibitem[Petre et al.(2002)]{petre44} Petre, R., Kuntz, K. D., \&
Shelton, R. L. 2002, ApJ, 579, 404
\bibitem[Reach \& Rho(1996)]{rr96} Reach, W. T., \& Rho, J.-H. 1996, A\& A, 315, L277
\bibitem[Reach \& Rho(1998)]{rr98} Reach, W. T., \& Rho, J.-H. 1998, ApJ, 507, L93
\bibitem[Reach \& Rho(1999)]{rr99} Reach, W. T., \& Rho, J.-H. 1999, ApJ, 511, 836
\bibitem[Reach \& Rho(2000)]{rr00} Reach, W. T., \& Rho, J.-H. 2000, ApJ, 544, 843
\bibitem[Reach et al.(2002)]{rr02} Reach, W. T., Rho, J.-H.,
Jarrett, T. H., \& Lagage, P.-O., 2002, ApJ, 564, 302
\bibitem[Rho \& Borkowski(2002)]{rho28} Rho, J., \& Borkowski, K. J. 2002, ApJ, 575, 201
\bibitem[Rho et al.(1994)]{rho44} Rho, J., Petre, R., Schlegel, E. M., Hester, J. J. 1994,
ApJ, 430, 757
\bibitem[Rho \& Petre(1998)]{RP98} Rho, J.-H., \& Petre, R. 1998, \apjl, 503, L167
\bibitem[Rho et al.(2001)]{rho443} Rho, J., Jarrett, T., Cutri, R., \& Reach, W. T.
2001, ApJ, 547, 885
\bibitem[Rho et al.(2003)]{rhocasa} Rho, J., Reynolds, S. P., Reach, W. T.,
Jarrett, T. H., Allen, G. E., \& Wilson, J. C.
2003, ApJ, 529, 299
\bibitem[Rowell et al.(2000)]{rowell} Rowell, G. P. et al. 2000, \aap, 359, 337
\bibitem[Sakamoto et al.(1994)]{sakamoto94} Sakamoto, S., Hayashi, M., 
Hasegawa, T., Handa, T., \& Oka, T. 1994, ApJ, 425, 641
\bibitem[Scalo(1986)]{imf} Scalo, J. M. 1985, Fund. Cosmic Phys., 11, 1
\bibitem[Schilke et al.(1997)]{shilke}
Schilke1, P., Walmsley, C. M., Pineau des For\^ets, G., \& Flower, D. R. 1997,
\aap, 321, 293
\bibitem[Seta et al.(1998)]{setaw44} Seta, M. et al. 1998, ApJ, 505, 286
\bibitem[Seta et al.(2004)]{seta04} Seta, M. Hasegawa, T., Sakamoto, S.,
Oka, T., Sawada, T., Insutsuka, S., Koyama, H., \& Hayashi, M. 
2004, AJ, 127, 1098
\bibitem[Seta(1995)]{setathesis} Seta, M. 1995, Ph. D. thesis, University of Tokyo
\bibitem[Shelton et al.(1999)]{shelton} Shelton, R. L., Cox, D. P., Maciejewski, 
W., Smith, R. K., Plewa, T., Pawl, A., \& R\'ozyczka, M. 1999, ApJ, 524, 192
\bibitem[Skrutskie(1999)]{twomassref} Skrutskie, M. F. 1999, in {\it Astrophysics with Infrared
Surveys: A Prelude to SIRTF}, eds. M. D. Bicay, C. A. Beichman, R. M. Cutri, \&
B. F. Madore (San Francisco: ASP), p. 185.
\bibitem[Spitzer(1978)]{spitzer} Spitzer, L. 1978, {\it Physical Processes in the 
Interstellar Medium} (New York: Wiley)
\bibitem[Van den Bergh, Marscher, \& Terzian(1973)]{vandenbergh}
Van den Bergh, S., Marscher, A. P., \& Terzian, Y. 1973, \apjs, 26, 19
\bibitem[Vel\'azquez et al.(2002)]{velazquez} Vel\'azquez, P. F.,
Dubner, G. M., Goss, W. M., \& Green, A. J. 2002, AJ, 124, 2151
\bibitem[Wardle and Yusef-Zadeh(2002)]{wardle} Wardle, M., \& Yusef-Zadeh, F.
2002, Science, 296, 2350
\bibitem[Wheeler(1981)]{wheeler81} Wheeler, J. C. 1981, Rep. Prog. Phys., 44, 6
\bibitem[White and Long(1991)]{whitelong} White, R. L., \& Long, K. S. 
1991, ApJ, 373, 543
\bibitem[Williams, Blitz, and Stark(1995)]{williams} Williams, J. P., Blitz, L.,
\& Stark, A. A. 1995, ApJ, 451, 252
\bibitem[Wilner, Reynolds, \& Moffett(1998)]{wilner} Wilner, D. J.,
Reynolds, S. P., \& Moffett, D. A. 1998, AJ, 115, 247
\bibitem[Wilson and Rood(1994)]{co1312_ratio}Wilson, T. L., \& Rood, R., 1994,
ARA\&A, 32, 191
\bibitem[Wolszczan,Cordes, and Dewey(1991)]{wolzscan} Wolszczan, A.,
Cordes, J. M., \& Dewey, R. J. 1991, ApJ, 372, L99
\bibitem[Woosley(1987)]{woosley} Woosley, S. E. 1987, in
{\it The origin and evolution of neutron stars}, eds. D.J. Helfand and 
J.-H. Huang (Dordrecht: Reidel), 255
\bibitem[Wootten(1977)]{wootten77} Wootten, H. A. 1977, ApJ, 216, 440 
\bibitem[Wootten(1981)]{wootten} Wootten, A. 1981, ApJ, 245, 105 
\bibitem[Yusef-Zadeh et al.(2003)]{yusefzadeh03a} Yusef-Zadeh, F.,
Wardle, M., \& Roberts, D. A. 2003, ApJ, 583, 267
\bibitem[Yusef-Zadeh et al.(2003)]{yusefzadeh03} Yusef-Zadeh, F.,
Wardle, M., Rho, J., \& Sakano, M. 2003, ApJ, 585, 319
\bibitem[Ziurys et al.(1989)]{ziurys}
Ziurys, L. M., Snell, R. L., \& Dickman, R. L. 1989, \apj, 341, 857
\end{thebibliography}
\end{document}